\documentclass[12pt]{article}
\usepackage{a4}
\usepackage{array}
\usepackage{calc}
\newlength{\adressabstand}
\input{diagrams}
\newarrow{auf}|---{->}
\newarrow{nach}----{->}
\newarrow{ident}33333
\newarrow{aufsurj}|---{->>}
\newarrow{nachsurj}----{->>}
\newlength{\CDhoehe}                 
\newlength{\CDgap}                 
\usepackage{ifthen,array}
\newcommand{\ams}{\usepackage{amsfonts,amssymb,amsmath}}

\ams
\allowdisplaybreaks[3]
\newlength{\textwidthorig}
\newlength{\oddsidemarginorig}
\newlength{\textheightorig}
\newlength{\topmarginorig}
\def\seitenlaengenabsolut#1 #2 #3 #4 {\setlength{\textwidth}{#1}
                                      \setlength{\oddsidemargin}{#2}
                                      \setlength{\textheight}{#3}
                                      \setlength{\topmargin}{#4}}
\def\seitenlaengenrelzustandard#1 #2 #3 #4 {\setlength{\textwidth}{\textwidthorig+#1}
                                            \setlength{\oddsidemargin}{\oddsidemarginorig+#2}
                                            \setlength{\textheight}{\textheightorig+#3}
                                            \setlength{\topmargin}{\topmarginorig+#4}}
\def\seitenlaengenrelzuvorher#1 #2 #3 #4 {\addtolength{\textwidth}{#1}
                                          \addtolength{\oddsidemargin}{#2}
                                          \addtolength{\textheight}{#3}
                                          \addtolength{\topmargin}{#4}}
\newcommand{\standardseite}{\seitenlaengenrelzuvorher2.2cm -0.8cm 1.8cm -1.5cm }   
\standardseite
\newcommand{\Wegdamit}[1]{}
\newcommand{\leerezeile}{\vspace{2ex}}
\newcommand{\einbett}{\hookrightarrow}   
\newcommand{\nach}{\longrightarrow}      
\newcommand{\txtnach}[1]{\xrightarrow{#1}}
\newcommand{\auf}{\longmapsto}           
\newcommand{\txtauf}[1]{\auf}            
\newcommand{\impliz}{\Longrightarrow}    
\newcommand{\aequ}{\Longleftrightarrow}  
 
\newcommand{\invimpliz}{\Longleftarrow}  
\newcommand{\iso}{\cong}                 
\newcommand{\ident}{\equiv}              
\newcommand{\teilmenge}{\subseteq}       
\newcommand{\obermenge}{\supseteq}       
\newcommand{\echteteilmenge}{\subset}    
\newcommand{\aeqrel}{\sim}               

\newcommand{\esex}{\exists}              
\newcommand{\kreuz}{\times}              

\newcommand{\einschr}[1]{\mid_{#1}}      
\newcommand{\dirprod}{\operatorname*{{\text{\Large$\boldsymbol{\kreuz}$}}}}
\newcommand{\betraganpass}[1]%
           {\left| #1 \right|}           
\newcommand{\betrag}[1]%
           {\betraganpass{#1}}           
\newcommand{\betragnichtanpass}[1]%
           {\mid #1 \mid}                
\newcommand{\norm}[1]%
           {\parallel #1 \parallel}      
\newcommand{\erww}[1]%
           {\langle #1 \rangle}          
\newcommand{\skalprod}[2]%
           {\langle #1,#2 \rangle}       
\newcommand{\quer}{\overline}            
\newcommand{\inv}[1]{\frac{1}{#1}}       
\newcommand{\im}{\text{im\;}}                          
\newcommand{\pr}{{\text{pr}}}                          
\newcommand{\id}{\:\text{id}}                          
\newcommand{\inter}{\text{int}\:}                      
\newcommand{\Ad}{{\text{Ad}}}                          
\newcommand{\elanz}{\#}                                
\newcommand{\Hom}{\text{Hom}}                          
\newcommand{\Maps}{\text{Maps}}                        
\newcommand{\field}[1]{\mathbb{#1}}                    
\newcommand{\N}{{\field{N}}}                           
\newcommand{\R}{{\field{R}}}                           
\newcommand{\Z}{{\field{Z}}}                           
\newcommand{\rnkl}[2]{\raisebox{-0.5ex}{$#1$}%
\raisebox{-0.2ex}{{\Large$\setminus$}}\,#2}            
\newcommand{\agb}{{\overline{{\cal A}/{\cal G}}}}      
\newcommand{\agbfact}[1][]{\text{$\agb/\!\aeqrel$}}    
\newcommand{\ag}{{\cal A}/{\cal G}}                    
\newcommand{\Ab}{{\overline{{\cal A}}}}                
\newcommand{\A}{{\cal A}}                              
\newcommand{\Gb}{{\overline{{\cal G}}}}                
\newcommand{\G}{{\cal G}}                              
\newcommand{\AbGb}{{\Ab/\Gb}}                          
\newcommand{\agsob}{{\A_{\text{Sob}}/\G_{\text{Sob}}}} 
\newcommand{\qa}{{\quer{A}}}                           
\newcommand{\qg}{{\quer{g}}}                           
\newcommand{\hg}{{\cal HG}}                            

\newcommand{\holgr}{{\mathbf H}}                       
\newcommand{\bz}{{\mathbf B}}                          
\newcommand{\Web}{{\text{Web}}}                        
\newcommand{\PT}{{\text{Type}}}                        
\newcommand{\eo}{{\mathbf E}}                          
\newcommand{\GR}{\Gamma}                               
\newcommand{\Ver}{\mathbf{V}}                          
\newcommand{\Edg}{\mathbf{E}}                          
\newcommand{\gross}[1]{{\boldsymbol #1}}               
\newcommand{\ga}{\gross{\alpha}}                       
\newcommand{\gb}{\gross{\beta}}                        
\newcommand{\pf}[1]{{\cal P}_{#1}}                     
\newcommand{\Pf}{{\cal P}}                             
\newcommand{\KG}[1]{\Pf_{#1}}                          
\newcommand{\LG}{{\mathbf{G}}}                         
\newcommand{\aeqrelzush}[1][]{\sim}                    
\newcommand{\Abiso}{{\cal I}}                          
\newcommand{\multi}{\text{m}}                          
\newcommand{\redisoeo}{\Psi_0}                         

\newcommand{\nklza}[1][]{\ifthenelse{\equal{#1}{}}     
                                    {\rnkl{Z(\holgr_\qa)}{\LG}}        
                                   {\rnkl{Z(\holgr_{#1})}{\LG}}}       
\newcommand{\nkla}[1][]{\ifthenelse{\equal{#1}{}}      
                                    {\rnkl{\bz(\qa)}{\Gb}}        
                                    {\rnkl{\bz(#1)}{\Gb}}}       
\newcommand{\wg}{\widetilde g}       %
\newcommand{\ListNullAbstaende}{\setlength{\topsep}{0pt}%
                                \setlength{\parskip}{0pt}%
                                \setlength{\partopsep}{0pt}%
                                \setlength{\itemsep}{0pt}%
                                \setlength{\parsep}{0pt}}
\newcommand{\ListNurAnstrichAbstand}{\setlength{\topsep}{0pt}%
                                     \setlength{\parskip}{0pt}%
                                     \setlength{\partopsep}{0pt}%
                                     \setlength{\parsep}{0pt}}
\newenvironment{StandardListe}[2]%
               {\begin{list}%
                      {#1}%
                      {\settowidth{\leftmargin}{M#1}%
                       \settowidth{\labelwidth}{#1}%
                       \settowidth{\labelsep}{M}%
                       #2%
                      }%
                }%
               {\end{list}}%
\newenvironment{EinfachListe}[1]%
               {\begin{StandardListe}{#1}{\ListNullAbstaende}}%
               {\end{StandardListe}}%
               {\begin{StandardListe}{#1}{\ListNurAnstrichAbstand}}%
               {\end{StandardListe}}%
\newcommand{\labelsatz}[1]{#1}
\newcounter{listennr}                      
\newlength{\hilfslaenge}
\newlength{\stdlabellaenge}
\newlength{\maximum}
\newcommand{\stdlabel}{}
\newcommand{\Maximum}{}
\newcommand{\iitem}[1][]{\ifthenelse{\equal{#1}{}}%
                           {\item \setlength{\hilfslaenge}{\stdlabellaenge}}%
                           {\item[\labelsatz{#1}\hfill]%
                            \settowidth{\hilfslaenge}{\labelsatz{#1}}}%
                         \ifthenelse{\lengthtest{\maximum < \hilfslaenge}}%
                           {\setlength{\maximum}{\hilfslaenge}%
                            \ifthenelse{\equal{#1}{}}%
                               {\renewcommand{\Maximum}{\stdlabel}}%
                               {\renewcommand{\Maximum}{#1}}}%
                           {}%
                      }      
\makeatletter
\newenvironment{AutoLabelLaengenListe}[2][]%
               {\begin{list}%
                      {\labelsatz{#1}\hfill}%
                      {\stepcounter{listennr}%
                       \settowidth{\leftmargin}{M\labelsatz{\ref{listnr\arabic{listennr}}}}%
                       \settowidth{\labelwidth}{\labelsatz{\ref{listnr\arabic{listennr}}}}%
                       \settowidth{\labelsep}{M}%
                       \settowidth{\stdlabellaenge}{\labelsatz{#1}}%
                       \renewcommand{\stdlabel}{#1}%
                       #2%
                       \renewcommand{\Maximum}{}%
                      }%
                }%
               {\renewcommand{\@currentlabel}{\Maximum}%
                \label{listnr\arabic{listennr}}%
                \end{list}%
                }%
\makeatother
\newenvironment{StandardEinrueckung}[2]%
               {\begin{list}%
                      {#1}%
                      {\settowidth{\leftmargin}{M#1}%
                       \settowidth{\labelwidth}{#1}%
                       \settowidth{\labelsep}{M}%
                       #2%
                      }%
                \item}%
               {\end{list}}%
\newenvironment{Einrueckungpur}[1]%
               {\begin{StandardEinrueckung}{#1}{\ListNullAbstaende}}%
               {\end{StandardEinrueckung}}%
\newenvironment{Einrueckung}[1]%
               {\begin{StandardEinrueckung}{#1}{\setlength{\parsep}{0pt}}}%
               {\end{StandardEinrueckung}}%
\newcommand{\EineZeileGleichung}[2][0.0ex]
           {
            
            \vspace{#1} 
            \noindent
            \hspace*{\fill}
            $\displaystyle{#2}$
            \hspace*{\fill}

            \vspace{#1} 
            
           }

\makeatletter
\newcommand{\EineNumZeileGleichung}[2][0.5ex]
           {
            
            \vspace{#1} 
            \noindent
            \stepcounter{equation}
            \renewcommand{\@currentlabel}{\arabic{equation}}%
            \phantom{(\arabic{equation})}\hspace*{\fill}
            $\displaystyle{#2}$
            \hspace*{\fill}
            (\arabic{equation})

            \vspace{#1} 
            
           }
\makeatother

\newlength{\abstaug}              
\newenvironment{AllgUnnumGleichung}[2][1.0ex]
               {
  
                \setlength{\abstaug}{#1}
                \vspace{\abstaug}
                \hspace*{\fill}
                $\begin{array}[t]{#2}
                }%
               {\end{array}$
                \hspace*{\fill}
  
                \vspace{\abstaug}

                }%
               {
  
                \setlength{\abstaug}{#1}
                \vspace{\abstaug}
                $\begin{tabular*}{\textwidth}[t]{#2}
                }%
               {\end{tabular*}$

                \vspace{\abstaug}

               }%
\newenvironment{StandardUnnumGleichung}[1][0ex]
               {\renewcommand{\s}{\\[#1] }%
                \begin{AllgUnnumGleichung}{>{\displaystyle}rc>{\displaystyle}l}}%
               {\end{AllgUnnumGleichung}}%
\newcommand{\erl}[1]{\hfill\mbox{\hspace*{1.5em}\small (#1)}}

\newcommand{\erllang}[2][0.5\textwidth]%
              {\hfill\hspace*{1.5em}%
               \begin{minipage}[t]{#1}{\small%
                          \begin{list}{(}{\ListNullAbstaende%
                                          \settowidth{\leftmargin}{(}%
                                          \settowidth{\labelwidth}{(}%
                                          \settowidth{\labelsep}{}%
                                         }%
                          \item#2)%
                          \end{list}}%
               \end{minipage}\\[-0.9ex]
              }%
\newcommand{\DefBemUmgeb}[1]
           {\newenvironment{#1}[1][]%
                           {\begin{Einrueckung}{{\bf #1}}%
                            \ifx##1\empty\else{{\bf ##1}
                            
                                                        }\fi%
                            }%
                           {\end{Einrueckung}}}
\newcommand{\DefSBemUmgeb}[2]
           {\newenvironment{#1}[1][]%
                           {\begin{Einrueckung}{{\bf #2}}%
                            \ifx##1\empty\else{{\bf ##1}
                            
                                                        }\fi%
                            }%
                           {\end{Einrueckung}}}
\makeatletter
\newcommand{\DefBspUmgeb}[3]
           {\newcounter{#2}[#3]%
            \newenvironment{#1}[1][]%
                           {\stepcounter{#2}%
                            \renewcommand{\ZaehlerMarke}{\arabic{#2}}%
                            \renewcommand{\Einzugsname}{{\bf #1 \ZaehlerMarke}}%
                            \begin{Einrueckung}{\Einzugsname}
                            \ifx##1\empty\else{{\bf ##1}\\}\fi%
                            \renewcommand{\@currentlabel}{\ZaehlerMarke}%
                            }%
                           {\end{Einrueckung}}}
\makeatother
\newcommand{\ZaehlerbisEbene}{section}
\newcommand{\Ebenea}{section}
\newcommand{\Ebeneb}{subsection}

\newcommand{\Abschnittnummer}{%
            \ifx\ZaehlerbisEbene\Ebenea{\arabic{section}}%
             \else{%
              \ifx\ZaehlerbisEbene\Ebeneb{\arabic{section}.\arabic{subsection}}%
               \else{\arabic{section}.\arabic{subsection}.\arabic{subsubsection}}%
              \fi}%
            \fi}     
\newcommand{\Abschnittnummerpunkt}{\Abschnittnummer.}     
\newcommand{\Einzugsname}{}
\newcommand{\ZaehlerMarke}{}
\makeatletter
\newcommand{\DefThmUmgeb}[3]
           {\newcounter{#1}[#3]%
            \newenvironment{#1}[1][]%
                           {\stepcounter{#2}%
                            \setcounter{#1}{\value{#2}}%
                            \renewcommand{\ZaehlerMarke}{\Abschnittnummerpunkt\arabic{#1}}%
                            \renewcommand{\Einzugsname}{{\bf #1 \ZaehlerMarke}}%
                            \begin{Einrueckung}{\Einzugsname}
                            \ifx##1\empty\else{{\bf ##1}
                            
                                                        }\fi%
                            \renewcommand{\@currentlabel}{\ZaehlerMarke}%
                            }%
                           {\end{Einrueckung}}}
\makeatother
\makeatletter
\newcommand{\DefSThmUmgeb}[4]
           {\newcounter{#1}[#3]%
            \newenvironment{#1}[1][]%
                           {\stepcounter{#2}%
                            \setcounter{#1}{\value{#2}}%
                            \renewcommand{\ZaehlerMarke}{\Abschnittnummerpunkt\arabic{#1}}%
                            \renewcommand{\Einzugsname}{{\bf #4 \ZaehlerMarke}}
                            \begin{Einrueckung}{\Einzugsname}
                            \ifx##1\empty\else{{\bf ##1}

                                                        }\fi%
                            \renewcommand{\@currentlabel}{\ZaehlerMarke}%
                            }%
                           {\end{Einrueckung}}}
\makeatother
\newenvironment{Beweis}[1][]%
               {\begin{Einrueckung}{{\bf Beweis}}%
                \ifx#1\empty\else{{\bf #1}

                                            }\fi%
                }%
               {\end{Einrueckung}%
                }%
\newenvironment{Proof}[1][]%
               {\begin{Einrueckung}{{\bf Proof}}%
                \ifx#1\empty\else{{\bf #1}

                                            }\fi%
                }%
               {\end{Einrueckung}%
                }%
               {\begin{Einrueckung}{{\bf \glqq Beweis\grqq}}%
                \ifx#1\empty\else{{\bf #1}
                
                                            }\fi%
                }%
               {\end{Einrueckung}%
                }%
               {\begin{Einrueckung}{{\bf Begr"undung}}%
                \ifx#1\empty\else{{\bf #1}
                
                                            }\fi%
                }%
               {\end{Einrueckung}%
                }%
\newenvironment{Hinrichtung}%
               {\begin{Einrueckungpur}{$\impliz$}}%
               {\end{Einrueckungpur}}%
\newenvironment{Rueckrichtung}%
               {\begin{Einrueckungpur}{$\invimpliz$}}%
               {\end{Einrueckungpur}}%
               {\begin{Einrueckungpur}{\glqq$\teilmenge$\grqq}}%
               {\end{Einrueckungpur}}%
               {\begin{Einrueckungpur}{\glqq$\obermenge$\grqq}}%
               {\end{Einrueckungpur}}%
               {\begin{Einrueckungpur}{"$\teilmenge$"}}%
               {\end{Einrueckungpur}}%
               {\begin{Einrueckungpur}{"$\obermenge$"}}%
               {\end{Einrueckungpur}}%
\newcommand{\qed}{\nopagebreak\hspace*{2em}\hspace*{\fill}{\bf qed}}
\newcommand{\ARabic}{\arabic}
\newcommand{\Nummerntypa}{\arabic}   
\newcommand{\Nummerntypb}{\alph}
\newcommand{\Nummerntypc}{\roman}
\newcommand{\Nummerntypd}{\Alph}

\newcommand{\Nra}{\Nummerntypa{Nummera}}            
\newcommand{\Nrb}{\Nummerntypb{Nummerb}}            
\newcommand{\Nrc}{\Nummerntypc{Nummerc}}                
\newcommand{\Nrd}{\Nummerntypd{Nummerd}}                
\newcommand{\ZeichenzuNrTyp}[1]%
           {\ifx#1\ARabic {.}\else{)}%
                  \fi}                              
\newcommand{\NrZeicha}{\ZeichenzuNrTyp{\Nummerntypa}}
\newcommand{\NrZeichb}{\ZeichenzuNrTyp{\Nummerntypb}}
\newcommand{\NrZeichc}{\ZeichenzuNrTyp{\Nummerntypc}}
\newcommand{\NrZeichd}{\ZeichenzuNrTyp{\Nummerntypd}}
\newcommand{\ListMarkea}%
           {\Nra\NrZeicha}
\newcommand{\ListMarkeb}%
           {\Nra\NrZeicha\Nrb\NrZeichb}
\newcommand{\ListMarkec}%
           {\Nra\NrZeicha\Nrb\NrZeichb\Nrc\NrZeichc}
\newcommand{\ListMarked}%
           {\Nra\NrZeicha\Nrb\NrZeichb\Nrc\NrZeichc\Nrd\NrZeichd}
\newcommand{\Anfangszeichen}{}
\newcommand{\Anfangspunkt}{}
\newcounter{Schachtelebene}
\newcounter{Hilfszaehler}
\newcommand{\Hilfsbefehl}{}
\newcommand{\Schachtelebene}{\alph{Schachtelebene}}
\makeatletter
\newenvironment{AllgNumerierteListe}[2][]
               {\addtocounter{Schachtelebene}{1}%
		\setcounter{Hilfszaehler}{#2}%
                \renewcommand{\Anfangszeichen}%
                             {\renewcommand{\Hilfsbefehl}{\csname Nummerntyp\Schachtelebene \endcsname}%
                              \Hilfsbefehl{Hilfszaehler}}%
                \renewcommand{\Anfangspunkt}%
                             {\csname NrZeich\Schachtelebene \endcsname}%
                \begin{list}%
                      {\stepcounter{Nummer\Schachtelebene}%
                       \csname Nr\Schachtelebene \endcsname
                       \csname NrZeich\Schachtelebene \endcsname
                       }%
                      {\settowidth{\leftmargin}{M\Anfangszeichen\Anfangspunkt}%
                       \settowidth{\labelwidth}{\Anfangszeichen\Anfangspunkt}%
                       \settowidth{\labelsep}{M}%
                       \setlength{\topsep}{0pt}%
                       \setlength{\parskip}{0pt}%
                       \setlength{\partopsep}{0pt}%
                       \setlength{\itemsep}{0pt}%
                       \setlength{\parsep}{0pt}%
                      }%
                \renewcommand{\@currentlabel}{\csname ListMarke\Schachtelebene \endcsname}%
                }%
               {\ifthenelse{\equal{}{}}{\setcounter{Nummer\Schachtelebene}{0}}{}
                \addtocounter{Schachtelebene}{-1}%
                \end{list}}
\makeatother
\newenvironment{NumerierteListe}[1]
               {\begin{AllgNumerierteListe}{#1}}
               {\end{AllgNumerierteListe}}
\newenvironment{WeiterNumerierteListe}[1]
               {\begin{AllgNumerierteListe}[Weiter]{#1}}
               {\end{AllgNumerierteListe}}

\newcommand{\UnnumAnfangszeichen}{}
\newcounter{UnnumSchachtelebene}
\newcommand{\UnnumSchachtelebene}{\alph{UnnumSchachtelebene}}
\makeatletter
\newenvironment{UnnumerierteListe}%
               {\addtocounter{UnnumSchachtelebene}{1}%
                \renewcommand{\UnnumAnfangszeichen}%
                             {\csname UnnumZeich\UnnumSchachtelebene \endcsname}%
                \begin{list}%
                      {\UnnumAnfangszeichen}%
                      {\settowidth{\leftmargin}{M\UnnumAnfangszeichen}%
                       \settowidth{\labelwidth}{\UnnumAnfangszeichen}%
                       \settowidth{\labelsep}{M}%
                       \setlength{\topsep}{0pt}%
                       \setlength{\parskip}{0pt}%
                       \setlength{\partopsep}{0pt}%
                       \setlength{\itemsep}{0pt}%
                       \setlength{\parsep}{0pt}%
                      }%
                }%
               {\addtocounter{UnnumSchachtelebene}{-1}%
                \end{list}}
\makeatother
\newlength{\fktdefhilfslaenge}
\newcommand{\fktdef}[5]
           {\hspace*{\fill}
            $\begin{array}[t]{cccc}%
            #1: & #2 & \nach & #3 \\    
                & #4 & \auf  & #5
            \end{array}$
            \settowidth{\fktdefhilfslaenge}{$#1$:}
            \hspace*{0.6 \fktdefhilfslaenge}  
            \hspace*{\fill}}
\newcommand{\fktdefpur}[5]
           {$\begin{array}[t]{cccc}%
            #1: & #2 & \nach & #3 \\    
                & #4 & \auf  & #5
            \end{array}$}
\newcommand{\fktdefabgesetzt}[5]
           {
           
            \hspace*{\fill}
            $\begin{array}[t]{cccc}%
            #1: & #2 & \nach & #3 \\    
                & #4 & \auf  & #5
            \end{array}$
            \settowidth{\fktdefhilfslaenge}{$#1$:}
            \hspace*{0.6 \fktdefhilfslaenge}  
            \hspace*{\fill}
            
            }
\newcommand{\sectioninh}[1]%
           {\section*{#1}%
            \addcontentsline{toc}{section}{#1}}
\newcommand{\anhang}%
           {\appendix
            \sectioninh{Anhang}
            \renewcommand{\Abschnittnummer}{%
                  \ifx\ZaehlerbisEbene\Ebenea{\Alph{section}}%
                  \else{%
                        \ifx\ZaehlerbisEbene\Ebeneb{\Alph{section}.\arabic{subsection}}%
                        \else{\Alph{section}.\arabic{subsection}.\arabic{subsubsection}}%
                        \fi}%
                  \fi}%
            \renewcommand{\Abschnittnummerpunkt}{\Abschnittnummer.}     
            }            
\newcommand{\anhangengl}%
           {\appendix
            \sectioninh{Appendix}
            \renewcommand{\Abschnittnummer}{%
                  \ifx\ZaehlerbisEbene\Ebenea{\Alph{section}}%
                  \else{%
                        \ifx\ZaehlerbisEbene\Ebeneb{\Alph{section}.\arabic{subsection}}%
                        \else{\Alph{section}.\arabic{subsection}.\arabic{subsubsection}}%
                        \fi}%
                  \fi}%
            \renewcommand{\Abschnittnummerpunkt}{\Abschnittnummer.}     
            }            
\newlength{\querfhilfsl}              

\newlength{\hll}
\newcommand{\cursorzurueck}[1]{
                               \settowidth{\hll}{$#1$}
                               \hspace*{-\hll}}

\newcommand{\bitteseitenumbr}{\pagebreak[4]}

\DefThmUmgeb{Theorem}{Theorem}{\ZaehlerbisEbene}
\DefThmUmgeb{Definition}{Definition}{\ZaehlerbisEbene}
\DefThmUmgeb{Satz}{Theorem}{\ZaehlerbisEbene}
\DefThmUmgeb{Proposition}{Theorem}{\ZaehlerbisEbene}
\DefThmUmgeb{Lemma}{Theorem}{\ZaehlerbisEbene}
\DefThmUmgeb{Folgerung}{Theorem}{\ZaehlerbisEbene}
\DefThmUmgeb{Corollary}{Theorem}{\ZaehlerbisEbene}
\DefThmUmgeb{Vorschrift}{Definition}{\ZaehlerbisEbene}
\DefThmUmgeb{Construction}{Definition}{\ZaehlerbisEbene}
\DefSThmUmgeb{FormSatz}{Theorem}{\ZaehlerbisEbene}{\glqq Satz\grqq} 
\DefThmUmgeb{Vermutung}{Theorem}{\ZaehlerbisEbene}
\DefBspUmgeb{Beispiel}{Beispiel}{subsubsection}
\DefBemUmgeb{Bemerkung}
\DefBemUmgeb{Remark}
\DefSBemUmgeb{OffeneFrage}{Offene Frage}
\newcommand{\bdf}{\begin{Definition}}
\newcommand{\edf}{\end{Definition}}
\newcommand{\bvorsch}{\begin{Vorschrift}}
\newcommand{\evorsch}{\end{Vorschrift}}
\newcommand{\bconst}{\begin{Construction}}
\newcommand{\econst}{\end{Construction}}
\newcommand{\bthm}{\begin{Theorem}}
\newcommand{\ethm}{\end{Theorem}}
\newcommand{\bsatz}{\begin{Satz}}
\newcommand{\esatz}{\end{Satz}}
\newcommand{\bprop}{\begin{Proposition}}
\newcommand{\eprop}{\end{Proposition}}
\newcommand{\blem}{\begin{Lemma}}
\newcommand{\elem}{\end{Lemma}}
\newcommand{\bfolg}{\begin{Folgerung}}
\newcommand{\efolg}{\end{Folgerung}}
\newcommand{\bcorr}{\begin{Corollary}}
\newcommand{\ecorr}{\end{Corollary}}
\newcommand{\bbew}{\begin{Beweis}}
\newcommand{\ebew}{\end{Beweis}}
\newcommand{\bpf}{\begin{Proof}}
\newcommand{\epf}{\end{Proof}}
\newcommand{\bwnum}{\begin{WeiterNumerierteListe}}
\newcommand{\ewnum}{\end{WeiterNumerierteListe}}
\newcommand{\bbem}{\begin{Bemerkung}}
\newcommand{\ebem}{\end{Bemerkung}}
\newcommand{\brem}{\begin{Remark}}
\newcommand{\erem}{\end{Remark}}
\newcommand{\bnum}{\begin{NumerierteListe}}
\newcommand{\enum}{\end{NumerierteListe}}
\newcommand{\bunum}{\begin{UnnumerierteListe}}
\newcommand{\eunum}{\end{UnnumerierteListe}}
\newcommand{\bbsp}{\begin{Beispiel}}
\newcommand{\ebsp}{\end{Beispiel}}
\newcommand{\bof}{\begin{OffeneFrage}}
\newcommand{\eof}{\end{OffeneFrage}}
\newcommand{\bgl}{\begin{StandardUnnumGleichung}}
\newcommand{\egl}{\end{StandardUnnumGleichung}}
\newcommand{\zgl}{\EineZeileGleichung}

\newcommand{\znumgl}{\EineNumZeileGleichung}
\newcommand{\berlgl}{\begin{StandardUnnumGleichung}}
\newcommand{\eerlgl}{\end{StandardUnnumGleichung}}
\newcommand{\beinrueck}{\begin{Einrueckungpur}} 
\newcommand{\eeinrueck}{\end{Einrueckungpur}}
\newcommand{\beinflist}{\begin{EinfachListe}} 
\newcommand{\eeinflist}{\end{EinfachListe}}
\newcommand{\beq}{\begin{equation}}
\newcommand{\eeq}{\end{equation}}
\newcommand{\bhin}{\begin{Hinrichtung}}
\newcommand{\ehin}{\end{Hinrichtung}}
\newcommand{\brueck}{\begin{Rueckrichtung}}
\newcommand{\erueck}{\end{Rueckrichtung}}
\newcommand{\bvl}{\begin{AutoLabelLaengenListe}{\ListNullAbstaende}}
\newcommand{\evl}{\end{AutoLabelLaengenListe}}
\newcommand{\df}[1]{{\bf #1}}
\newcommand{\zglnum}[2]{\znumgl{#1\label{#2}}}
\begin{document}
\title{Gauge Orbit Types for Generalized Connections}
\author{Christian Fleischhack\thanks{e-mail: 
            Christian.Fleischhack@itp.uni-leipzig.de {\it or}    
            Christian.Fleischhack@mis.mpg.de} \\   
        \\
        \begin{minipage}{0.43\textwidth}
        \begin{center}
        {\normalsize\em Mathematisches Institut}\\[\adressabstand]
        {\normalsize\em Universit\"at Leipzig}\\[\adressabstand]
        {\normalsize\em Augustusplatz 10/11}\\[\adressabstand]
        {\normalsize\em 04109 Leipzig, Germany}\\
        \end{center}
        \end{minipage}
        \begin{minipage}{0.43\textwidth}
        \begin{center}
        {\normalsize\em Institut f\"ur Theoretische Physik}\\[\adressabstand]
        {\normalsize\em Universit\"at Leipzig}\\[\adressabstand]
        {\normalsize\em Augustusplatz 10/11}\\[\adressabstand]
        {\normalsize\em 04109 Leipzig, Germany}\\
        \end{center}
        \end{minipage} \\[-10\adressabstand]
        {\normalsize\em Max-Planck-Institut f\"ur Mathematik in den
                        Naturwissenschaften}\\[\adressabstand]
        {\normalsize\em Inselstra\ss e 22-26}\\[\adressabstand]
        {\normalsize\em 04103 Leipzig, Germany}}
\date{January 5, 2000}
\maketitle
\begin{abstract}
Different versions for defining Ashtekar's generalized connections 
are investigated depending on
the chosen smoothness category for the paths and graphs
-- the label set for the projective limit. Our definition
covers the analytic case as well as the case of webs.

Then the orbit types of the generalized connections are determined
for compact structure groups.
The stabilizer of a connection is homeomorphic to the 
holonomy centralizer, i.e. the centralizer
of its holonomy group, and the homeomorphism class of 
the gauge orbit is completely determined by the 
holonomy centralizer. Furthermore,
the stabilizers of two connections are conjugate in the gauge group
if and only if their holonomy centralizers are conjugate in
the structure group.
Finally, the gauge orbit type of a connection is defined to be
the conjugacy class of its holonomy centralizer equivalently to the 
standard definition via stabilizers.
\end{abstract}
\newpage
\section{Introduction}
For a few decades the quantization of Yang-Mills theories has been 
investigated extensively. 
One of the most important approaches uses functional integration. 
Here one quantizes a classical theory by introducing an appropriate measure
on its configuration space.
In gauge theories this space is given by $\ag$.
Here, originally, $\G$ denoted the set of all (smooth) 
gauge transforms acting on the space $\A$ of all 
(smooth) connections.
That is why a lot of the efforts has been focussed 
on clarifying the structure of $\ag$. 
One typical property of $\ag$ is that there
do not exist global gauge fixings, i.e. smooth
sections in $\A\nach\ag$, -- the so-called Gribov problem.
Other problems are caused by 
the very difficult structure of $\ag$: $\ag$ is non-linear, infinite-dimensional 
and it is usually not a manifold.
Thus, results concerning $\ag$ are quite scarce up to now.
But, should one restrict oneself to the case of smooth connections?
Since in a quantization process smoothness is usually
lost anyway, it is quite clear that one has to admit also non-smooth connections.
This way, about 20 years ago, several authors started the consideration
of Sobolev connections. For basic results we refer, e.g., to \cite{f1}.
By now, the structure, in particular, of the generic stratum of $\agsob$
is quite well-understood.
Nevertheless, measure theory did not become easier.
Concerning that point, first convincing successes have been gained through 
the introduction of generalized connections by Ashtekar and
Isham \cite{a72}.
Here one drops completely the "differential" conditions like
smoothness or Sobolev integrability and works with the
algebraic structure of the space of connections only.
The main idea is as follows. A (smooth) connection is uniquely
determined by its parallel transports, i.e., by a (smooth \cite{d20}) 
homomorphism from the groupoid of paths to the structure group $\LG$.
A generalized connection is now simply such a homomorphism, but 
without the smoothness condition.
Analogously, a generalized gauge transform $\qg\in\Gb$
is now a (usually non-smooth)
map from the base manifold $M$ to $\LG$, and it acts purely
algebraically on the space $\Ab$ of generalized connections.
One of the main advantages of $\Ab$ is that it is (for compact $\LG$)
compact and it possesses a natural kinematical measure, 
the induced Haar measure \cite{a48}. 
Now, the perhaps most important question is how the standard smooth
and the new Ashtekar
theory are related to each other -- mathematically and physically.
The first very nice answer was the statement that 
$\A$ is dense in $\Ab$ \cite{e8}.
This result is usually expected when one quantizes a theory.
Then it has been proven that for the two-dimensional
pure Yang-Mills theory the Wilson loop expectation values are in
fact the same in the classical as well as in the Ashtekar framework
\cite{a6,paper1}. Now, we are going to investigate the action 
of the generalized gauge transforms on the space of generalized connections 
in comparison 
with its counterpart in the Sobolev case 
described in detail in \cite{f11}.

The present paper is the first in a series of three papers. 

In the first part of this paper we will give a quite 
detailed introduction into the algebraic and topological 
definitions and properties of 
$\Ab$, $\Gb$ and $\AbGb$. Here we closely follow
Ashtekar and Lewandowski \cite{a30,a28} as well as 
Marolf and Mour\~ao \cite{a42}. The most important difference
to their definitions is that we do not restrict the paths to be
(piecewise) analytic or smooth. For our purpose it is sufficient to fix
a category of smoothness from the beginning. This is $C^r$, where 
$r$ can be any positive natural number, $\infty$ (smooth case) or
$\omega$ (analytical case). We can also consider the corresponding 
cases $C^{r,+}$ of paths that are (piecewise) immersions.
We will show that in a certain sense the case
$(\omega,+)$ corresponds to the loop structures introduced by Ashtekar
and Lewandowski \cite{a48} and the case $(\infty,+)$ corresponds to
the webs introduced by Baez and Sawin \cite{d3}.

Now, the line of our papers ramifies. One branch, described in
the second paper \cite{paper3} of our short series, investigates 
properties of the space $\Ab$ itself. 
There we will give a construction method
for new connections. Then, as a main result, 
we will show that an induced Haar measure $d\mu_0$ can be defined
for arbitrary smoothness conditions. For this, we introduce the
notion of a hyph that generalizes the notion of a web and a graph.
We show that the paths of a hyph are holonomically independent
and that the set of all hyphs is directed. These two properties
yield the well-definedness of $d\mu_0$.

The other branch is followed in the second part of the present paper.
It is devoted to the type of the gauge orbit.
In the general theory of transformation groups the type of an orbit (or, more
precisely, an element of an orbit) is
defined by the conjugacy class of 
its stabilizer (see, e.g., \cite{Bredon}).
Here, we will derive the explicit
form of the stabilizer for every generalized connection.
As we will see, the stabilizer of a connection is homeomorphic to 
the centralizer of its holonomy group, hence a finite-dimensional Lie
group. Since stabilizers are conjugate in $\Gb$ if and only if these
centralizers are conjugated in $\LG$, the type of an orbit is uniquely determined
by a certain equivalence class of a Howe subgroup of the structure group $\LG$.
(A Howe subgroup of $\LG$ is a subgroup that can be written as the
centralizer of some subset $V\teilmenge\LG$.)

In the final paper \cite{paper4} of this short series  
we reunite the two branches. There we will see how the results
of Kondracki and Rogulski \cite{f11} can be extended from the Sobolev
framework to the generalized case (for compact $\LG$).
We will prove that there is a slice theorem 
for the action of $\Gb$ on $\Ab$.
This means that for every connection $\qa\in\Ab$ 
there is an open and $\Gb$-invariant neighbourhood 
that can be retracted equivariantly to the orbit $\qa\circ\Gb$.
Moreover, we prove that the space $\AbGb$ is topologically
regularly stratified. But, two results for generalized connections
go beyond those for Sobolev ones. First, we can explicitly derive the
set of all gauge orbit types. 
This was not known until now for the Sobolev case. However, recently,
Rudolph, Schmidt and Volobuev \cite{f14} solved this problem for
all $SU(n)$-bundles over two-, three- and four-dimensional
manifolds.
We show that in the Ashtekar framework
(the conjugacy class of) every Howe subgroup of $\LG$
occurs as a gauge orbit type. 
Second, we prove that the generic stratum, i.e. the set of all
connections whose holonomy centralizers equal the center of $\LG$, has the
induced Haar measure $1$.

\leerezeile

In the following, $M$ is always a connected and at least two-dimensional 
$C^r$-manifold with $r\in\N^+\cup\{\infty\}\cup\{\omega\}$ being arbitrary, but 
fixed. Furthermore, $m$ is an, as well, arbitrary, but fixed point in $M$ and
$\LG$ is a Lie group.

\section{Paths}
\label{sect:paths}
In the classical approach a connection can be described
by the corresponding parallel transports along paths in the base
manifold. But, not every assignment of group elements to the paths
yields a connection. On the one hand, this map has to be a homomorphism, 
i.e., products of paths have to lead to products of the parallel transports,
and on the other hand, it has to depend in a certain sense continuously
on the paths. Moreover, additional topological obstructions may
occur. In the Ashtekar approach, however, the second (and the third) condition
is dropped. A connection is now simply a homomorphism from the
set of paths to the structure group $\LG$.

Up to now, it is not clear, whether there is an "optimal" definition
for the structure of the groupoid $\Pf$ of paths. The first version was
given by Ashtekar and Lewandowski \cite{a48}. They used piecewise analytical
paths. The advantage of this approach was that any finite set of paths
forms a finite graph. Hence for two finite graphs there is always a third
graph containing both of them, i.e. the set of all graphs forms a directed
set. Using this it is easy to prove
independence theorems for loops and to define then a natural measure on $\Ab$
and $\agb$. But, the restriction to analyticity seems a little bit 
unsatisfactory. Since one has desired from the very beginning
to use $\Ab$ for describing quantum gravity, one comes into troubles
with the diffeomorphism invariance of this theory: After applying
a diffeomorphism a path need no longer be analytical.

That is why Baez and Sawin \cite{d3} introduced so-called
webs and tassels built by only smooth paths fulfilling certain conditions.
Any graph can be written as a web and for any finite number of webs there
is a web containing all of them. So the directedness of the label set
for the definition of $\Ab$ remaines valid, and, consequently, 
one can generalize the construction of the 
natural induced Haar measure and lots of things more. 

In this paper we will introduce another definition for paths.
Our definition will have the advantage that it does not depend explicitly
on the chosen smoothness category labelled by 
$r\in\N^+\cup\{\infty\}\cup\{\omega\}$. Moreover,
it does not matter, whether we demand the paths to be piecewise
immersions (cases $C^{r,+}$) or not. Therefore, in what follows
suppose that we have fixed the parameter $r$ from the very beginning.
Furthermore, we decide now whether we additionally demand
the paths to be piecewise immersions or not.
Nevertheless, we write always simply $C^r$.
\subsection{General Case}
In this subsection we consider all smoothness categories on one stroke.
\bdf
\label{def:path}
A \df{path} is a piecewise
$C^r$-map $\gamma: [0,1] \nach M$.\footnote{If we consider piecewise
immersed paths, we have to additionally define all $\gamma: [0,1] \nach M$
that are piecewise constant, i.e. 
$\gamma\einschr{[\tau_1,\tau_2]} = \{x\}$ for some $x\in M$, or immersive
to be a path.}

The \df{initial point} is $\gamma(0)$ and the \df{terminal point} $\gamma(1)$.

Two paths $\gamma_1$ und $\gamma_2$ can be multiplied
iff the terminal point of $\gamma_1$ and the initial point of $\gamma_2$ 
coincide. Then the product is given by  
\zgl{\gamma_1\gamma_2(t) := \begin{cases} 
                        \gamma_1(2t) & \text{ for } 0\leq t\leq\inv2 \\
                        \gamma_2(2t-1) & \text{ for } \inv2\leq t\leq 1 
                        \end{cases}.}

A path $\gamma$ is called \df{trivial} iff 
$\im\gamma\ident\gamma([0,1])$ is a single point.
\edf
An important idea of the Ashtekar program is the assumption
that the total information about the continuum theory is encoded in the 
set of all subtheories on finite lattices. Thus we need the definition
of paths and graphs.
The set of all paths is hard to manage. That is why we restrict ourselves
to special paths.
\bdf
\label{def:simplepath}
\label{def:finitepath}
\label{def:pathequiv}
\bunum
\item
A path $\gamma$ has \df{no self-intersections}
iff from $\gamma(\tau_1) = \gamma(\tau_2)$ follows that
\bunum
\item
$\tau_1 = \tau_2$ or 
\item
$\tau_1 = 0$ and $\tau_2 = 1$ or
\item
$\tau_1 = 1$ and $\tau_2 = 0$.
\eunum
\item
A path $\gamma'$ is called \df{subpath} of a path $\gamma$ iff
there is an affine non-decreasing map $\phi:[0,1]\to[0,1]$ with
$\gamma' = \gamma\circ\phi$. Iff additionally $\phi(0)=0$ (or
$\phi(1)=1$), $\gamma'$ is called \df{initial path} (or \df{terminal path})
of $\gamma$.

We define $\gamma^{t,+}(\tau):=\gamma(t+\tau(1-t))$ for all $t\in[0,1)$
and $\gamma^{t,-}(\tau):=\gamma(\tau t)$ for all $t\in(0,1]$ to be the 
outgoing and incoming subpath of $\gamma$ in $t$, respectively.

If $\gamma$ is a path without self-intersections then set
$\gamma^{x,\pm}:=\gamma^{t,\pm}$ for all $x\in\im\gamma$ 
where $t$ fulfills $\gamma(t) = x$. (We choose $t=0$ in the $+$-case
if $x=\gamma(0)$. Analogously for $t=1$.)
\item
A (finite) \df{graph} $\GR$ 
is a (finite) union of paths $e_i$ without self-intersections
and of isolated points $v_j$. The elements of  
$\Ver(\GR):=\bigcup_i\{e_i(0), e_i(1)\} \cup \bigcup_j\{v_j\}$ 
are called \df{vertices}, that of 
$\Edg(\GR):=\bigcup_i\{e_i\}$ \df{edges}. A graph $\GR$ is called \df{connected}
iff $\Ver(\GR)\cup\bigcup_{e\in\Edg(\GR)}\im e$ is connected.
\item
A \df{path in a graph} $\GR$ is a path in $M$, that 
equals a product of edges in $\GR$ and
trivial paths (with values in $\Ver(\GR)$), respectively, 
whereas the product of two consecutive paths has to exist.

A path $\gamma$ in $M$ is called \df{simple} iff there is a finite 
graph $\GR$ such that $\gamma$ is a path in $\GR$.
\item
A path $\gamma$ in $M$ is called \df{finite} iff $\gamma$
is up to
the parametrization\footnote{Two paths $\gamma_1$ and $\gamma_2$
are equal up to the parametrization iff there is a  
bijective $\Pi:[0,1]\nach [0,1]$ with $\Pi(0)=0$ and 
$\gamma_2 = \gamma_1\circ\Pi$ such that $\Pi$ and $\Pi^{-1}$ are $C^r$.}
equal to a finite product of simple 
paths.
\item
Two finite paths $\gamma_1$ and $\gamma_2$ are called \df{equivalent} iff
there is a finite sequence of finite paths $\delta_i$ with
$\delta_0 = \gamma_1$ and $\delta_n = \gamma_2$ such that for 
all $i = 1, \ldots, n$
\bunum
\item
$\delta_i$ and $\delta_{i-1}$ coincide up to the parametrization or
\item
$\delta_i$ arises from $\delta_{i-1}$ by inserting\footnote{This means,
there is a $\tau\in[0,1]$ and a finite path $\delta$ such that

$\delta_i(t) = \begin{cases}
                \delta_{i-1}(\inv2 t) 
                    & \text{ for $\phantom{\inv2\tau{}+{}}\:0\leq t\leq\inv2\tau$} \\
                \delta(4(t-\inv2\tau)) 
                    & \text{ for $\inv2\tau\phantom{{}+{}\inv2}
                                     \leq t\leq\inv2\tau+\inv4$} \\
                \delta(4(\inv2\tau+\inv2-t)) 
                    & \text{ for $\inv2 \tau+\inv4\leq t\leq\inv2\tau+\inv2$}\\
                \delta_{i-1}(\inv2 t-\inv2) 
                    & \text{ for $\inv2\tau+\inv2\leq t
                                     \leq\:\phantom{\inv2\tau{}+{}}1$}
                \end{cases}$.
          
In the following, we denote by a retracing of a path $\gamma$ a subpath 
of the form $\delta\delta^{-1}$ with a certain finite $\delta$.}
a retracing
or
\item
$\delta_{i-1}$ arises from $\delta_i$ by inserting a retracing.
\eunum
\item
The set of all classes of finite paths is denoted by $\Pf$, that of paths
in $\GR$ by $\KG\GR$.
Furthermore, we write $\pf{xy}$ for the set of all classes of finite paths
from $x$ to $y$.
The set of all classes of finite paths having base point $m$ 
forms the \df{hoop group} $\hg\ident\pf{mm}$.
\eunum
\edf
We have immediately from the definitions
\bprop
The multiplication on $\Pf$ induced by the multiplication of paths is
well-defined and generates a groupoid structure on 
$\Pf$.\footnote{This means, roughly speaking, $\Pf$ possesses
all properties of a group: associativity, existence of unit elements and
of the inverse. But, the product need not be defined for all paths.}

The hoop group $\hg$ is a subgroup of $\Pf$.
\eprop
\brem
\bnum{5}
\item
One can define an analogous equivalence relation
on the set of paths in a fixed graph: Two paths would be
"$\GR$-equivalent", iff they arise from each other by reparametrizations
or by inserting or cancelling of retracings contained in $\GR$.

Obviously, two paths in $\GR$ are equivalent, if they are $\GR$-equivalent.
On the other hand, one can also prove that two paths contained in $\GR$
are already $\GR$-equivalent if they are equivalent.

Consequently, we can identify $\KG\GR$ and the set of all 
$\GR$-equivalence classes of paths in $\GR$. In other words:
$\KG\GR$ is the groupoid that is
generated freely by the set of all edges of $\GR$.
\item
In what follows we usually say instead of "finite connected graph" simply
"graph" and instead of "finite path" only "path".
Moreover, by a path we always mean 
-- if not explicitly the converse is said -- 
an equivalence class of paths.
\item
Finally, we identify two graphs if the (corresponding)
edges are equivalent. Since edges are per def. free of retracings,
this simply means that the edges are equal up to the parametrization.
\item
Note that the paths $\gamma_1(\tau) := \tau$ and 
$\gamma_2(\tau) := \tau^2$ in $\R(\teilmenge\R^n)$ 
are not equivalent. This comes from the
fact that $\Pi:\tau\nach\tau^2$ is $C^r$, but $\Pi^{-1}:\tau\nach\sqrt\tau$
is not. (As well, it is not possible to transform $\gamma_1$ into $\gamma_2$
successively inserting or deleting retracings as in Definition 
\ref{def:pathequiv}.) Furthermore, one sees that
$\gamma_1\circ\gamma_2^{-1}$ is an example for a path with retracings
that is not equivalent to a path without. 
\item
If we restricted ourselves to piecewise analytical paths, i.e. the 
smoothness category $(\omega,+)$
from the very beginning,
every path would be finite. \cite{a48} 
\enum
\erem
The main assumption quoted above suggests the usage of finite graphs as an
index set for the subtheories. But, these theories are not "independent".
Roughly speaking, a subtheory defined on a smaller lattice arises by 
projecting the theory defined on the bigger lattice.
\bdf
Let $\GR_1$ and $\GR_2$ be two graphs. $\GR_1$ is  
\df{smaller or equal} $\GR_2$ ($\GR_1\leq\GR_2$)
iff each edge of $\GR_1$ is
(up to the parametrization) a product of edges of $\GR_2$ and 
the vertex sets fulfill $\Ver(\GR_1)\teilmenge\Ver(\GR_2)$.
\edf
Obviously, $\leq$ is a partial ordering.
\subsection{Immersive Case}
In the case of piecewise immersed paths we can define another 
equivalence relation for finite paths. Here we use the fact that
any piecewise immersed path can be parametrized proportionally
to the arc length:
\bdf
We shortly call a path a \df{pal-path} iff it is parametrized
{\em p}\/roportionally to the {\em a}\/rc {\em l}\/ength.

Two finite paths $\gamma_1$ and $\gamma_2$ are called \df{equivalent} iff
there is a finite sequence of finite paths $\delta_i$ with
$\delta_0 = \gamma_1$ and $\delta_n = \gamma_2$ such that for 
all $i = 1, \ldots, n$
\bunum
\item
$\delta_i$ and $\delta_{i-1}$ coincide
when parametrized proportionally
to the arc length\footnote{This 
definition seems to require a certain Riemannian structure
on $M$. But, on the one hand, each manifold can be given a Riemannian
structure. On the other hand, the definition of equivalence does not
depend on the chosen Riemannian metric:
if two paths coincide w.r.t. to the arc length to the first metric then
they obviously coincide w.r.t. to the arc length of the other metric.
Thus, this definition is indeed completely determined by the manifold
structure of $M$.} or
\item
$\delta_i$ arises from $\delta_{i-1}$ by inserting a retracing
or
\item
$\delta_{i-1}$ arises from $\delta_i$ by inserting a retracing.
\eunum
\edf
\blem
\bnum{4}
\item
Two finite paths $\gamma_1$ and $\gamma_2$ 
are equivalent if they can be obtained from each
other by a reparametrization.
\item
Each nontrivial finite path is equivalent to a pal-path without retracings.
\enum
\elem
\bpf
\bnum{4}
\item
Clear.
\item
We prove this inductively on the number $n$ of simple paths $\gamma_i$
that the finite path $\gamma$ is decomposed into. We will even prove that 
$\gamma$ is equivalent to a pal-path $\gamma'$ that can be decomposed 
(up to the parametrization) into $n'\leq n$ simple paths and 
that has no retracings.

For $n=1$ we have
nothing to prove. Thus, let $n\geq 2$. First free 
$\gamma_0:=\prod_{i=1}^{n-1}\gamma_i$ off the retracings using the induction 
hypothesis. We get a pal-path $\gamma'_0\aeqrel\prod_{i=1}^{n'-1}\gamma'_i$
with the desired properties and $n'\leq n$.
Denote by $\gamma'$ the pal-path corresponding 
to $\gamma'_0\gamma_n$. Obviously,
$\gamma'\aeqrel\gamma$. Suppose, $\gamma'$ is not free of retracings.
Let $\delta\delta^{-1}$ be a retracing. 
Then a part of the retracing $\delta\delta^{-1}$ has to be 
in $\gamma_n$. Since $\gamma_n$ is simple (and w.l.o.g. non-trivial), the
terminal point of $\delta$ cannot be in $\inter\gamma_n$. Since by assumption
$\gamma'_0$ is free of retracings, the terminal point has to be
the initial point of $\gamma_n$, and thus $\delta^{-1}$ is (if necessary, after
an appropriate [affine] reparametrization) an initial path of $\gamma_n$.
Assume now $\delta$ to be maximal, i.e., any $\delta$ "containing"
terminal path $\delta'$ of $\gamma'_0$ that yields a retracing in $\gamma'$
is equal to $\delta$.\footnote{Such a $\delta$ exists:
Assume that every pal-subpath $\delta_\tau^{-1}$
of $\gamma_n$ corresponding to the parameter interval
$[0,\tau]$ with $\tau<T$ yields a retracing. (Such a $T$ exists, because
there exists some retracing.) By the continuity of every path and the fact
that the paths arising here and so all their subpaths are pal, also 
$\delta_T^{-1}$ has to yield a retracing. Consequently, there is a maximal
$T$.}
Now, cancel out the retracing: If $\delta$ is not a (genuine) 
subpath of $\gamma'_{n'-1}$
(i.e., "exceeds" or equals it),
define $\gamma'_n$ to be the "remaining" part of $\gamma_n$ 
"outside" $(\gamma'_{n'-1})^{-1}$; then 
$\gamma'':=\bigl(\prod_{i=1}^{n'-2}\gamma'_i\bigr)\circ\gamma'_n\aeqrel\gamma'$
consists of at most $n'-2+1< n$ finite paths. The induction hypothesis gives 
the assertion.
Suppose now that $\delta$ is a (genuine) 
subpath of $\gamma'_{n'-1}$. Then define the pal-path
$\gamma''$ by 
$\bigl(\prod_{i=1}^{n'-2}\gamma'_i\bigr)\gamma''_{n'-1}\circ\gamma'_n$,
where $\gamma'_n$ denotes the "remaining" part of 
$\gamma_n$ outside of $\delta^{-1}$ and $\gamma''_{n'-1}$ that of
$\gamma'_{n'-1}$ outside of $\delta$. By the maximality of $\delta$,
$\gamma''$ contains no retracings.
$\gamma''\aeqrel\gamma'\aeqrel\gamma$ yields the assertion.
\qed
\enum
\epf
Most of the constructions in the following as well as most of those in the 
subsequent papers \cite{paper3,paper4} do actually not depend on the 
choice of the equivalence relation for the paths. But, the second one can
only be used for piecewise immersed paths. Therefore,
in what follows, we will use the general equivalence relation 
given in the last subsection.
\section{Gauge Theory on the Lattice}
\label{abschn:gittereth}
In this section we will transfer the lattice gauge theory given
by Ashtekar and Lewandowski \cite{a30,a28} to our case.
The algebraic definitions for the connections, gauge transforms and
the action of the latter ones follow these authors closely.
In the last two subsections we will state some
assertions mainly on the basic properties of the action of the
gauge transforms and the projections onto smaller graphs.

\subsection{Algebraic Definition}
We use the standard definition: Globally connections are parallel transports,
i.e. $\LG$-valued homomorphisms of paths in $M$,
and gauge transforms are $\LG$-valued functions over $M$. 
The lattice versions now come from restricting the domain of definition
to edges and vertices in a graph.
\bdf
Let $\GR$ be a graph. We define

$\Ab_\GR := \Hom(\KG\GR,\LG)$ $\ldots$ set of all connections on $\GR$ and

$\Gb_\GR := \Maps(\Ver(\GR),\LG)$ $\ldots$ set of all gauge transforms
on $\GR$.

Here, $\Hom(\KG{\GR},\LG)$ denotes the set of all homomorphisms from the
groupoid $\KG\GR$ freely generated by the edges of $\GR$ into the structure 
group and $\Maps(\Ver(\GR),\LG)$ the set of all maps from the set of all
vertices of $\GR$ into the structure group.
\edf
In the classical case the action of a gauge transform on a connection
can be described by the corresponding action on the parallel transports:
\zgl{h_A(\gamma) \auf g^{-1}_{\gamma(0)} h_A(\gamma) g_{\gamma(1)}.}
By simply restricting onto the lattice we receive the action 
of $\Gb_\GR$ on $\Ab_\GR$ by  
\fktdefabgesetzt{\Theta_\GR}{\Ab_\GR\kreuz\Gb_\GR}{\Ab_\GR}
                {(h_\GR,g_\GR)}{h_\GR\circ g_\GR}
with $h_\GR\circ g_\GR(\gamma) := 
             g_\GR(\gamma(0))^{-1} \: h_\GR(\gamma) \: g_\GR(\gamma(1))$
for all paths $\gamma$ in $\GR$.
\bdf
For each graph $\GR$ we define 

$\agb_\GR := \Ab_\GR/\Gb_\GR$ $\ldots$ set of all equivalence classes
of connections in $\GR$.
\edf
\subsection{Topological Definition}
It is obvious that the groupoid $\KG\GR$ is always freely generated by the 
edges $e_i$ of $\GR$. Hence, the set $\Ab_\GR=\Hom(\KG\GR,\LG)$
can be identified via $h\auf \bigl(h(e_1),\ldots,h(e_{\elanz\Edg(\GR)})\bigr)$ with
$\LG^{\elanz\Edg(\GR)}$ and can so be given a natural topology. 
Analogously, we use that naturally
$\Gb_\GR = \Maps(\Ver(\GR),\LG)$ can be identified via 
$g\auf(g(x))_{x\in\Ver(\LG)}$ with $\LG^{\elanz\Ver(\GR)}$. So $\Gb_\GR$ 
is by means of the pointwise multiplication a topological
group. We have immediately
\bprop
For all graphs $\GR$ the action 
$\Theta_\GR:\Ab_\GR\kreuz\Gb_\GR\nach\Ab_\GR$ is continuous.
\eprop
\bpf
$\Theta_\GR$ as a map from 
$\LG^{\elanz\Edg(\GR)}\kreuz\LG^{\elanz\Ver(\GR)}$
to $\LG^{\elanz\Edg(\GR)}$ is a concatenation of multiplications, 
hence continuous.
\qed
\epf
\bcorr
$\agb_\GR=\Ab_\GR/\Gb_\GR$ is a Hausdorff space. $\agb_\GR$ is
compact for compact $\LG$.
\ecorr
It is well-known that connections are dual to paths and equivalence classes
of connections are dual to closed paths. This is again confirmed by
\bprop
\label{agbGristHom}
$\agb_\GR$ is isomorphic to
$\Hom(\hg_{x,\GR},\LG)/\Ad$, hence isomorphic to $\LG^{\dim\pi_1(\GR)}/\Ad$,
for each graph $\GR$ and for each vertex $x$ in $\GR$.

Here $\hg_{x,\GR}$ is the set of all (classes of) path(s) in $\GR$ starting and
ending in $x$, and $\pi_1(\GR)$ is the fundamental group of $\GR$.
\eprop
\bpf
Define
\fktdef{J}{\agb_\GR}{\Hom(\hg_{x,\GR},\LG)/\Ad.}
          {[h]}{[h\einschr{\hg_{x,\GR}}]_\Ad}
\bunum
\item
$J$ is well-defined.

If $h'=h''\circ g$, then $h'(\alpha) = g(x)^{-1} h''(\alpha) g(x)$ for all
$\alpha\in\hg_{x,\GR}$, i.e. 
$h'\einschr{\hg_{x,\GR}} = h''\einschr{\hg_{x,\GR}} \circ_\Ad g(x)$.
\item
$J$ is injective.

Let $J(h') = J(h'')$, i.e., let there exist a $g\in\LG$ such that
$h'(\alpha) = g^{-1} h''(\alpha) g$ for all $\alpha\in\hg_{x,\GR}$.
Choose for all vertices $y\neq x$ a path
$\gamma_y$ from $x$ to $y$, set $\gamma_x := 1$ and set
$g(y):= h''(\gamma_y)^{-1}\:g\:h'(\gamma_y)$ for all $y$. Now, 
$h' = h'' \circ (g(y))_{y\in\Ver(\GR)}$ is clear.
\item
$J$ is surjective.

Let $[h]\in\Hom(\hg_{x,\GR},\LG)/\Ad$ be given.
Choose an $h\in [h]$ and as above for all vertices $y$ a path $\gamma_y$ and
some $g_y\in\LG$. 
For each $\gamma\in\KG\GR$ set
$h_0(\gamma) := g_{\gamma(0)}^{-1} \:
                h(\gamma_{\gamma(0)}\gamma\gamma_{\gamma(1)}^{-1})\:
                g_{\gamma(1)}$.
We have $J(h_0) = [h]$.              
\eunum
Since $\hg_{x,\GR}$ is isomorphic to $\pi_1(\GR)$, hence a free group with 
$\dim\pi_1(\GR)$ generators \cite{paper1,a48}, we have 
$\agb_\GR \iso \LG^{\dim\pi_1(\GR)}/\Ad$.
\qed
\epf

\subsection{Relations between the Lattice Theories}
If one constructs a global theory from its subtheories one has to
guarantee that these subtheories are "consistent".
This means, e.g.,
that the projection of a connection onto a smaller graph has to be already 
defined by its projection onto a bigger graph. So we need
projections onto the subtheories induced by the partial ordering 
on the set of graphs.
\bdf
Let $\GR_1\leq\GR_2$.

We define
\fktdefabgesetzt{\pi_{\GR_1}^{\GR_2}}{\Ab_{\GR_2}}{\Ab_{\GR_1},}
                            {h}{h\einschr{\KG{\GR_1}}}
\fktdefabgesetzt{\pi_{\GR_1}^{\GR_2}}{\Gb_{\GR_2}}{\Gb_{\GR_1}}
                            {g}{g\einschr{\Ver(\GR_1)}}
and
\fktdefabgesetzt{\pi_{\GR_1}^{\GR_2}}{\agb_{\GR_2}}{\agb_{\GR_1}.}
                            {[h]}{[h\einschr{\KG{\GR_1}}]}
\edf
We denote all the three maps by one and the same symbol because it should
be clear in the following what map is meant.

Obviously, from $h' = h'' \circ g$ on $\GR_2$ follows
$h'\einschr{\KG{\GR_1}} = h''\einschr{\KG{\GR_1}} \circ g\einschr{\Ver(\GR_1)}$
on $\GR_1$, i.e. $\pi_{\GR_1}^{\GR_2}$ is well-defined. Furthermore,
we have
\bprop
Let $\GR_1\leq\GR_2\leq\GR_3$. Then
$\pi_{\GR_1}^{\GR_2}\pi_{\GR_2}^{\GR_3} = \pi_{\GR_1}^{\GR_3}$.
\eprop
Finally, we write down the projections by operations on the structure group 
in order to see topological properties.

Let again $\GR_1\leq\GR_2$. First we decompose each edge
$e_i$ of $\GR_1$ into edges $f_j$ of $\GR_2$: 
$e_i = \prod_{k_i=1}^{K_i} f_{j(i,k_i)}^{\epsilon(i,k_i)}$.
With this we get for the map between the connections
($n := \elanz\Edg(\GR_1)$)
\fktdefabgesetzt{\pi_{\GR_1}^{\GR_2}}
                {\LG^{\elanz\Edg(\GR_2)}}{\LG^{\elanz\Edg(\GR_1)}.}
                {\Bigl(g_1,\ldots,g_{\elanz\Edg(\GR_2)}\Bigr)}
            {\Bigl(\prod_{k_1=1}^{K_1} g_{j(1,k_1)}^{\epsilon(1,k_1)},
                  \ldots,
                  \prod_{k_{n}=1}^{K_{n}} g_{j({n},k_{n})}^{\epsilon({n},k_{n})}
                  \Bigr)}
On the level of gauge transforms the description is very easy:
$\pi_{\GR_1}^{\GR_2}$ projects $(g_v)_{v\in\Ver(\GR_2)}$
onto those elements belonging to vertices in $\GR_1$.
For classes of connections an analogous formula as for connections
holds: First choose two free generating systems 
$\ga$ and $\gb$
of $\hg_{x_1,\GR_1}$ and $\hg_{x_2,\GR_2}$, respectively, and then
a path $\gamma$ from $x_2$ to $x_1$ in the bigger graph $\GR_2$. Thus we get
decompositions
$\alpha_i = \gamma^{-1} 
            \bigl(\prod_{k_i=1}^{K_i} \beta_{j(i,k_i)}^{\epsilon(i,k_i)}\bigr) 
            \gamma$.
Hence, ($n_i := \dim\pi_1(\GR_i)$)
\fktdefabgesetzt{\pi_{\GR_1}^{\GR_2}}
                {\LG^{n_2}/\Ad}{\LG^{n_1}/\Ad.}
                {\Bigl[g_1,\ldots,g_{n_2}\Bigr]_\Ad}
            {\Bigl[\prod_{k_1=1}^{K_1} g_{j(1,k_1)}^{\epsilon(1,k_1)},
                  \ldots,
     \prod_{k_{n_1}=1}^{K_{n_1}} g_{j({n_1},k_{n_1})}^{\epsilon({n_1},k_{n_1})}
                  \Bigr]_\Ad}
\bprop
\label{satz:pi=stet,off,surj}
$\pi_{\GR_1}^{\GR_2}$ is continuous, open and surjective.
\eprop
\bpf
The surjectivity is clear for all three cases.

The continuity is trivial for the first two cases and follows in the third
because the projections $\LG^n\nach \LG^n/\Ad$ are open, continuous 
and surjective (see \cite{Bredon}) 
and the map from $\LG^{n_2}$ to $\LG^{n_1}$ 
corresponding to $\pi_{\GR_1}^{\GR_2}$ is obviously continuous.

The openness follows immediately in the case of gauge transforms
because projections onto factors of a direct product are open anyway.
In the case of connections one additionally needs the openness of the
multiplication in $\LG$: Each edge in $\GR_1$ is a product of edges 
in $\GR_2$, i.e., after possibly renumbering we have 
$e_i = f_{i,1}\cdots f_{i,K_i}$. Thus, 
$\pi_{\GR_1}^{\GR_2}(g_{1,1},\ldots,g_{n,K_n},\ldots) = 
     (g_{1,1}\cdots g_{1,K_1},\ldots,g_{n,1}\cdots g_{n,K_n})$. 
Let now $W$ be open in $\Ab_{\GR_2} = \LG^{\elanz\Edg(\GR_2)}$. Then
$W$ is a union of sets of the form
$W_{1,1}\kreuz\cdots\kreuz W_{n,K_n}\kreuz\cdots$, i.e.,
$\pi_{\GR_1}^{\GR_2}(W)$ is a union of sets of the form
$(W_{1,1} \cdots W_{1,K_1})\kreuz \cdots \kreuz (W_{n,1}\cdots W_{n,K_n})$. 
But these are open, i.e., $\pi_{\GR_1}^{\GR_2}$ is open.
The openness of $\pi_{\GR_1}^{\GR_2}:\agb_{\GR_2}\nach\agb_{\GR_1}$ follows
now because the map $\pi_{\GR_1}^{\GR_2}:\Ab_{\GR_2}\nach\Ab_{\GR_1}$ is open
and the projections $\Ab_\GR\nach\agb_\GR$ are continuous, open and surjective.
\qed
\epf

\section{Continuum Gauge Theory}
For completeness in the first subsection we will briefly quote the 
definitions of $\Ab$, $\Gb$ and $\agb$ from \cite{a30} and in the second
we summarize the most important facts about these spaces.
In the last two subsections we will first investigate the topological
properties of the action of $\Gb$ on $\Ab$ and of the projections
onto the lattice gauge theories and then prove that the connections etc.
are algebraically described exactly in the same form both for our
definition of paths and for that of Ashtekar and Lewandowski \cite{a48}.

\subsection{Definition of $\Ab$, $\Gb$ and $\agb$}
\label{uabschn:def_Ab+Gb+agb}
By means of the continuity of the projections
$\pi_{\GR_1}^{\GR_2}$ the spaces $(\Ab_\GR)_\GR$, $(\Gb_\GR)_\GR$ 
and $(\agb_\GR)_\GR$ are projective systems of topological spaces. 
This leads to the crucial \cite{a30}
\bitteseitenumbr
\bdf[Generalized Gauge Theories]
\label{def:continuumtheory}
\bunum
\item
$\Ab := \varprojlim_\GR \: \Ab_\GR$ is the space of \df{generalized
connections}.

The elements of $\Ab$ are usually denoted by $\qa$ or $h_\qa$.
\item
$\Gb := \varprojlim_\GR \: \Gb_\GR$
is the space of \df{generalized gauge transforms}.

The elements of $\Gb$ are usually denoted by $\qg$.
\item
$\agb := \varprojlim_\GR \: \agb_\GR$ is the space of \df{generalized
equivalence classes of connections}.
\eunum
\edf
Explicitly this means
\zgl{\Ab = \{(h_\GR)_\GR \in \dirprod_\GR \: \Ab_\GR \mid 
            \pi_{\GR_1}^{\GR_2} \: h_{\GR_2} = h_{\GR_1} \text{ for all }
            \GR_1\leq\GR_2\},}
\znumgl{\Gb = \{(g_\GR)_\GR \in
            \dirprod_\GR \: \Gb_\GR \mid 
            \pi_{\GR_1}^{\GR_2} \: g_{\GR_2} = g_{\GR_1} \text{ for all }
            \GR_1\leq\GR_2\}\label{kompbedgb}}
as well as
\zgl{\agb = \{([h_\GR])_\GR \in \dirprod_\GR \: \agb_\GR \mid 
            \pi_{\GR_1}^{\GR_2} \: [h_{\GR_2}] = [h_{\GR_1}] \text{ for all }
            \GR_1\leq\GR_2\}.}
We denote 
\fktdefabgesetzt{\pi_\GR}{\Ab}{\Ab_\GR,}{(h_{\GR'})_{\GR'}}{h_\GR}
\fktdefabgesetzt{\pi_\GR}{\Gb}{\Gb_\GR}
                 {(g_{\GR'})_{\GR'}}{g_\GR}                         
and
\fktdefabgesetzt{\pi_\GR}{\agb}{\agb_\GR.}{([h_{\GR'}])_{\GR'}}{[h_\GR]}
\subsection{Topological Characterization of $\Ab$, $\Gb$ and $\agb$}
We have \cite{a30,Kelley}
\bthm
\label{thm:allg(ab,gb,agb)}
\bnum{4}
\item
$\Ab$, $\Gb$ and $\agb$ are completely regular Hausdorff spaces
and, for compact $\LG$, compact.
\item
For every principle fibre bundle over $M$ with structure group $\LG$
the regular connections (gauge transforms, equivalence classes of connections) 
are also generalized connections 
(gauge transforms, equivalence classes of generalized connections).
This means the maps $\A\nach\Ab$, $\G\nach\Gb$ and $\ag\nach\AbGb$ are 
embeddings.
\item
\label{spez_kritstetprojlim}
Let $X$ be a topological space.

A map $f:X\nach \Ab$ is continuous iff
$\pi_\GR \circ f : X\nach \Ab_\GR \ident \LG^{\elanz\Edg(\GR)}$ is
continuous for all graphs $\GR$.

The analogous assertion holds for maps from $X$ to $\Gb$ and $\agb$,
respectively, as well.
\item
$\pi_\GR$ is continuous
for all graphs $\GR$.
\item
$\Gb$ is a topological group.
\enum
\ethm
We shall postpone the discussion 
whether the space $\A$ is dense in $\Ab$ or not for several reasons.
This, in fact, depends crucially on the chosen smoothness category 
and equivalence relation for the paths. It should be clear that -- provided
$\gamma_1(\tau):=\tau$ and $\gamma_2(\tau):=\tau^2$ 
are seen to be non-equivalent -- the denseness is unlikely: 
No classical smooth connection $A$ can distinguish
between these paths. So we will discuss this a bit more in detail in the
subsequent paper \cite{paper3}. As well, we will show
there that $\pi_\GR$ is also open and surjective. But all that 
requires some technical efforts that are absolutely not necessary
for the actual goal of this paper -- the determination of the gauge orbit 
types.
\bpf
\bnum{3}
\item
The property of being compact, Hausdorff or completely regular is maintained
by forming product spaces and by the transition to closed subsets.
Thus the assertion follows from the corresponding properties of the 
structure group $\LG$.
\item
The embedding property follows from Giles' reconstruction theorem
\cite{e14} and \cite{a72}. 
\item
See, e.g., \cite{Kelley}.
\item
Since $\id:\Ab\nach\Ab$ etc. is continuous, this follows from 
the facts just proven.
\item
The multiplication on $\Gb$ is defined by
$(g_\GR)_\GR \circ (g'_\GR)_\GR = (g_\GR \circ g'_\GR)_\GR$.
With this $\Gb$ is a group with unit $(e_\GR)_\GR$ and inverse
$\bigl((g_\GR)_\GR\bigr)^{-1} = (g_\GR^{-1})_\GR$. The multiplication
$\multi:\Gb\kreuz\Gb\nach\Gb$ is continuous due to the 
continuity criterion above: 
$\pi_\GR\circ\multi = \multi_\GR\circ(\pi_\GR\kreuz\pi_\GR)$ is continuous
for all $\GR$, because the multiplication $\multi_\GR$ on $\Gb_\GR$
is continuous.
\qed
\enum
\epf
\subsection{Action of Gauge Transforms on Connections}
Because of the consistency of the actions of $\Gb_\GR$ on $\Ab_\GR$ 
one can also define an action of $\Gb$ on $\Ab$. One simply sets 
\cite{a30}
\fktdefabgesetzt{\Theta}{\Ab\kreuz\Gb}{\Ab.}
                {\bigl((h_\GR)_\GR,(g_\GR)_\GR\bigr)}{(h_\GR \circ g_\GR)_\GR}
\bthm
\bnum{3}
\item
The action $\Theta$ of $\Gb$ on $\Ab$ is continuous.
\item
The maps 

\fktdef{\qa}{\Gb}{\Ab}{\qg}{\qa\circ\qg} and
\fktdef{\qg}{\Ab}{\Ab}{\qa}{\qa\circ\qg}

are continuous.
\item
The canonical projection $\pi_\AbGb:\Ab\nach\AbGb$ is continuous and open
and for compact $\LG$ also closed and proper.
\item
The map \fktdef{\pi_\GR}{\AbGb}{\Ab_\GR/\Gb_\GR}
                        {[(h_{\GR'})_{\GR'}]}{[h_\GR]}
is well-defined and continuous.
\enum
\ethm
\bpf
\bnum{3}
\item
$\pi_\GR\circ\Theta = \Theta_\GR\circ(\pi_\GR\kreuz\pi_\GR) : 
  \Ab\kreuz\Gb\nach\Ab_\GR$ as a concatenation of continuous maps 
on the right-hand side is continuous for any graph $\GR$.
By the continuity criterion for maps to $\Ab$ 
in Theorem \ref{thm:allg(ab,gb,agb)}, $\Theta$ is continuous.
\item
Follows from the continuity of $\Theta$.
\item
Follows because $\Theta$ is a continuous
action of a (compact) topological group $\Gb$ on the Hausdorff space $\Ab$.
\cite{Bredon}
\item
$\pi_\GR$ is well-defined. Namely, let $\qa' = \qa\circ\qg$,
i.e. $(h'_{\GR'})_{\GR'}= (h_{\GR'} \circ g_{\GR'})_{\GR'}$, thus
$h'_{\GR'} = h_{\GR'} \circ g_{\GR'}$ for all graphs $\GR'$.
Then $[h'_\GR] = [h_\GR]$.

The continuity of $\pi_\GR:\AbGb\nach\Ab_\GR/\Gb_\GR$ 
follows from the continuity of  
$\pi_\GR:\Ab\nach\Ab_\GR$ and $\pi_{\Ab_\GR/\Gb_\GR}$ as well as 
from the continuity criterion for the quotient topology because the diagram
\begin{center}
\vspace*{\CDgap}
\begin{minipage}{8cm}
\begin{diagram}[labelstyle=\scriptstyle,height=\CDhoehe,l>=3em]
\Ab         & \relax\rnachsurj^{\pi_\AbGb}             & \AbGb \\
\relax\dnach^{\pi_\GR} &                               & \relax\dnach^{\pi_\GR} \\ 
\Ab_\GR     & \relax\rnachsurj^{\pi_{\Ab_\GR/\Gb_\GR}} & \Ab_\GR/\Gb_\GR
\end{diagram}
\end{minipage}
\end{center}
is commutative.
\qed
\enum
\epf
We note that for a compact structure group $\LG$ and for
analytic paths $\AbGb$ and $\agb$ are even homeomorphic (cf. \cite{a30,a28}).
\subsection{Algebraic Characterization of $\Ab$, $\Gb$ and $\AbGb$}
In this subsection we will show that our choice of the definition of paths
leads to the same results as the definitions in \cite{a48} do.  
\bthm
\label{thm:algchar(Ab+Gb+agb)}
\bnum{4}
\item
We have $\Ab \iso \Hom(\Pf,\LG).$\footnote{This justifies the notation
$h_\qa$ for a connection $\qa$.}

Here, $\Hom(\Pf,\LG)$ is the set of all maps $h:\Pf\nach\LG$, that fulfill
$h(\gamma_1\gamma_2) = h(\gamma_1) h(\gamma_2)$ for all 
multipliable paths $\gamma_1,\gamma_2\in\Pf$.
\item
We have $\Gb\iso\dirprod_{x\in M} \LG\ident\Maps(M,\LG)$.

The isomorphism is even a homeomorphism of topological groups.
\item
The action of gauge transforms on the connections is given by
\znumgl{h_{\qa\circ\qg}(\gamma) := 
        g^{-1}_{\gamma(0)} \: h_\qa(\gamma) \: g_{\gamma(1)} 
        \text{ for all } \gamma\in\Pf.} 
$h_\qa:\Pf\nach\LG$ is the homomorphism corresponding to $\qa\in\Ab$ and
$g_x$ the component of the gauge transform $\qg\in\Gb$ in $x$. 
\item
We have $\AbGb \iso \Hom(\hg,\LG)/\Ad$.

Here, $\Hom(\hg,\LG)$ is the set of all homomorphisms $h:\hg\nach\LG$.
\enum
\ethm
\bpf
\bnum{4}
\item
Define \fktdef{I}{\Hom(\Pf,\LG)}{\Ab.}{h}{(h\einschr{\KG\GR})_\GR}

\bunum
\item
$I$ is injective.

From $h_1\neq h_2$ follows the existence of a $\gamma\in\Pf$
with $h_1(\gamma) \neq h_2(\gamma)$. Since $\gamma$ equals $\prod\gamma_i$ 
with appropriate simple $\gamma_i$, we have
$\prod h_1(\gamma_i) \neq \prod h_2(\gamma_i)$, hence
$h_1(\gamma_i) \neq h_2(\gamma_i)$ for some $\gamma_i$.
Choose a finite graph $\GR$ such that $\gamma_i$ is a path in $\GR$. 
Here we have  
$     h_1\einschr{\KG\GR}(\gamma_i) = h_1(\gamma_i) 
 \neq h_2(\gamma_i) = h_2\einschr{\KG\GR}(\gamma_i)$, i.e.
$I(h_1)\neq I(h_2)$.
\item
$I$ is surjective. 

Let $(h_\GR)_\GR$ be given. 
We consider first not classes of paths, but the paths itself.
Construct for any simple $\gamma\in\Pf$ a graph $\GR$ containing
$\gamma$. 
Define
$h(\gamma) := h_\GR(\gamma)$. 
For general $\gamma\in\Pf$ define
$h(\gamma) := \prod h(\gamma_i)$ according to some decomposition of
$\gamma$ into simple paths $\gamma_i$.

This construction is well-defined: First one easily realizes that it is
independent of the decomposition of $\gamma$ into finite paths (thus also of
the parametrization).\footnote{Namely, let $\prod \gamma'_i$ 
and $\prod \gamma''_j$ be two decompositions of $\gamma$. 
The terminal points of $\gamma'_i$ and $\gamma''_j$ correspond to certain
values of the parameters $\tau'_i$ and $\tau'_j$, respectively, of
the path $\gamma$.
Order these values to a sequence $(\tau_k)$ and construct
a decomposition of $\gamma$ into simple paths $\delta_k$ such that
$\delta_k$ corresponds to the segment $\gamma\einschr{[\tau_k,\tau_{k+1}]}$.
Now, on the one hand, $\gamma$ equals up to the parametrization 
$\prod \delta_k$, but, on the other hand, each $\gamma'_i$ and
$\gamma''_j$ equals up to the parametrization a product
$\delta_\kappa\circ\delta_{\kappa+1}\circ\cdots\circ\delta_\lambda$
with certain $\kappa,\lambda$. 

Now let $\widetilde\GR'_i$ be that graph w.r.t. that $\gamma'_i$ is simple.
Construct hereof the graph $\GR'_i$ by
inserting the terminal points of all the $\gamma''_j$ 
as vertices. Finally, let $\GR_k$ be the graph spanned by $\delta_k$. 
Thus, $\GR_k,\widetilde\GR'_i\leq\GR'_i$, and we have
\bgl
h(\gamma'_i) 
 & = & h_{\widetilde\GR'_i}(\gamma'_i) \\
 & = & h_{\GR'_i}(\gamma'_i) \\
 & = & h_{\GR'_i}(\delta_\kappa\circ\delta_{\kappa+1}\circ\cdots
                  \circ\delta_\lambda) \\
 & = & h_{\GR'_i}(\delta_\kappa) \: h_{\GR'_i}(\delta_{\kappa+1}) \: \cdots \:
       h_{\GR'_i}(\delta_\lambda) \\
 & = & h_{\GR_\kappa}(\delta_\kappa) \: h_{\GR_{\kappa+1}}(\delta_{\kappa+1}) 
       \: \cdots \:
       h_{\GR_\lambda}(\delta_\lambda).
\egl
Using the analogous relation for $\gamma_j''$ we have
$\prod h(\gamma'_i) = \prod h_{\GR_k}(\delta_k) 
                    = \prod h(\gamma''_j)$.
Thus, $h(\gamma)$ does not depend on the decomposition.}
Hence obviously, $h$ is a homomorphism. Thus, also 
$h(\gamma'\delta\delta^{-1}\gamma'') = h(\gamma'\gamma'')$ etc., i.e.,
$h$ is constant on equivalence classes of paths.
Consequently, $h:\Pf\nach\LG$ is a well-defined
homomorphism with $I(h) = (h_\GR)_\GR$.
\eunum
\item
Set \fktdef{I}{\Maps(M,\LG)}{\Gb.}
              {(g_x)_{x\in M}}{\bigl((g_x)_{x\in\Ver(\GR)}\bigr)_\GR}

Obviously, $I$ is bijective and a group homomorphism.

The topology on $\Maps(M,\LG)=\dirprod_{x\in M} \LG$ is generated
by the preimages $\pi_y^{-1}(U)$ of open $U\teilmenge\LG$, by which 
$\pi_y:(g_x)_{x\in M}\auf g_y$ is continuous.
Hence, 
$\pi_\GR\circ I = \pi_{v_1} \kreuz \cdots \kreuz \pi_{v_{\elanz\Ver(\GR)}}$
is continuous for all $\GR$, i.e., $I$ is continuous.
Due to the continuity criterion for maps into product spaces,
also $I^{-1}$ is continuous because
for all $y$ the map $\pi_y\circ I^{-1} = \pi_\GR$ ($\GR$ consists only of the
vertex $y$) is continuous.
\item
This follows immediately from the preceding steps.
\item
Use the map
\fktdef{J}{\AbGb}{\Hom(\hg,\LG)/\Ad}{[h]}{[h\einschr{\hg}]_\Ad}
and repeat the steps of the proof of Proposition \ref{agbGristHom}.
\qed
\enum
\epf
In the following we will usually write a gauge transform in the form
$\qg = (g_x)_{x\in M}$.
Furthermore we have again by the continuity criterion for maps into
product spaces
\bcorr
\label{folg:stetkritGbx}
Let $X$ be a topological space.

A map $f:X\nach \Gb$ is continuous iff
$\pi_x \circ f : X\nach \LG$ is continuous for all
$x\in M$.

$\pi_x$ is continuous for all $x\in X$.
\ecorr
\brem
If we work in the $(\omega,+)$-category for the paths, i.e., we only
consider piecewise analytical graphs, all the definitions and results 
coincide completely with those of Ashtekar and Lewandowski in
\cite{a48,a30,a28}.
\erem

\section{Graphs vs. Webs}
In this section we will compare the consequences of our definition of
paths to that of webs \cite{d3,d17,e46}. Within this section we only
consider the smooth category\footnote{Remember, the $+$ means
that all the paths are piecewise immersions.} $(\infty,+)$
for paths. Note that, within this section, 
a path is simply a piecewise immersive and $C^\infty$-map 
from $[0,1]$ to $M$, i.e. it is not an equivalence class. 
But it is still finite as before.

Let us briefly quote the basic properties of webs. A web consists of a
finite number of so-called tassels. 
A tassel $T$ with base point $p\in M$ is a finite, ordered set of curves $c_i$
(piecewise immersive
smooth maps\footnote{Thus the notion of a curve coincides with our notion of a
general, usually non-finite path.} from $[0,1]$ to $M$)
that fulfills certain properties: 
\bnum{4}
\item
$c_i(0) = p$ for all $i$ (common initial point).
\item
$c_i$ is an embedding (in particular, has no self-intersections).
\item
There is a positive constant $k_i\in\R$ for each $i$ such that 
$c_i(t) = c_j(s)$ implies $k_i t = k_j s$ (consistent parametrization).
\item
Define $\PT(x):=\{i\in I\mid x\in\im\:c_i\}$ for all $x\in X$.
Then, for all $J\echteteilmenge I$ the set 
$\PT^{-1}(\{J\})$  
is empty or has $p$ as an accumulation point.
\enum
Thus, in our notation, each $c_i$ is a simple path.
A web is now a finite collection of tassels such that 
no path of one tassel contains the base point of another tassel.
The following theorem on curves proven by Baez and Sawin \cite{d3} will
be crucial:
\bthm
\label{lem:baez+sawin}
Given a finite set $C$ of curves. Then there is a web $w$, such that
every curve $c\in C$ is equivalent to a finite product of paths
$\gamma\in w$ and their inverses.
\ethm 
This, namely, leads immediately to the following
\bprop
\label{prop:curve=finite_path}
Every curve
is equivalent to a finite path.
\eprop
Thus, our restriction to finite paths is actually no restriction.
\bpf 
Let there be given an arbitrary curve $\gamma:[a,b]\nach M$. 
By the preceding theorem $\gamma$ depends on some web $W$, i.e., 
there is a family of curves $c_i$ being simple paths such that
$\gamma$ equals 
(modulo equivalence, i.e. up to reparametrizations, cf. \cite{d3})
a finite product of the curves $c_i$ and their inverses. By Definition
\ref{def:finitepath}, $\gamma$ is finite.
\qed
\epf
This means, roughly speaking, the sets of paths the connections are
based on are the same for the webs and our case $(\infty,+)$.
But this yields the equality of our definition of $\Ab$ and 
that of Baez and Sawin.
\bthm 
Suppose $\LG$ to be compact and semi-simple.

Then $\Ab_\Web$ and $\Ab_{(\infty,+)}$, i.e. the 
spaces of generalized connections defined by webs \cite{d3} and
by Definition \ref{def:continuumtheory}, respectively, are homeomorphic.
\ethm
\bpf
Using the proposition above we see analogously to the proof 
of Theorem \ref{thm:algchar(Ab+Gb+agb)} that 
\fktdefabgesetzt{I_\Web}{\Hom(\Pf,\LG)}{\Ab_\Web}{h}{(h\einschr{w})_w}
is a bijection. Thus, 
$\Abiso:=I_\Web\circ I^{-1}:\Ab_{(\infty,+)}\nach\Ab_\Web$ is
a bijection, too. We are left with the proof that $\Abiso$ is a homeomorphism.
For this it is sufficient to prove that each element of a subbase of the 
one topology has an open image in the other topology. Possible subbases
for $\Ab_{(\infty,+)}$ and $\Ab_\Web$ are the families of all sets
of the type $\pi_\GR^{-1}(W_\GR)$ and $\pi_w^{-1}(W_w)$, respectively.
Hereby, $w$ is a web and $W_w\teilmenge\LG^k$, $k$ being the number of paths
in $w$, open\footnote{This is the point where we need the semi-simplicity and
compactness of $\LG$, because only for these assumptions it is proven
\cite{e46} up to now 
that the projection 
$\pi_w\einschr\A:\Ab_\Web\obermenge\A\nach\LG^k$ is surjective, i.e.
$\A_w = \LG^k$.
Otherwise, it would be possible that $\pi_w(\A)$ is a non-open
Lie subgroup of $\LG^k$. So the sets $\pi_w^{-1}(W_w)$ do no
longer create a subbase.\label{foot:web=infty+}}; 
$\GR$ is a graph and $W_\GR\teilmenge\LG^{\elanz\Edg(\GR)}$ an element
of a certain subbase, e.g., a set of type 
$W_\GR = W_1\kreuz\cdots\kreuz W_{\elanz\Edg(\GR)}$ with open $W_i\in\LG$.
Thus, we can take as a subbase for $\Ab_{(\infty,+)}$ simply 
all sets $\pi_c^{-1}(W)$ where $c$ is a simple path, i.e. a graph, and
$W\teilmenge\LG$ is open. Since every web is a collection of 
a finite number of simple paths, we get completely analogously
that the family of all $\pi_c^{-1}(W)$ is a subbase for $\Ab_\Web$.
The only difference here is that $c$ has to be simple with different
initial and terminal point. We are therefore left with the proof that
$\Abiso(\pi_c^{-1}(W))$ is open in $\Ab_\Web$ for all simple, closed paths $c$ and
all open $W$, which is, however, quite easy. Decompose $c$ into two
paths $c_1$ and $c_2$ (with different initial and terminal points)
which span the graph $\GR$. Then
$\Abiso(\pi_c^{-1}(W))=\Abiso(\pi_\GR^{-1}((\pi_c^{\GR})^{-1}(W))$. 
By the continuity 
of $\pi_c^{\GR}$ the set $(\pi_c^{\GR})^{-1}(W)$ is open in $\LG^2$, i.e.
a union of sets of the type $W_1\kreuz W_2$, but
$\Abiso(\pi_\GR^{-1}(W_1\kreuz W_2)$ is open as discussed above.
\qed
\epf
\brem
We note that the homeomorphy of $\Ab_\Web$ and $\Ab_{(\infty,+)}$
remains valid also for arbitrary Lie groups $\LG$. But, for this proof
we need the surjectivity of $\pi_w$ for all webs as 
mentioned in Footnote \ref{foot:web=infty+}. This, on the other hand,
will be discussed in a subsequent paper \cite{paper3}.
\erem

\section{Determination of the Gauge Orbit Types}
Now we come to the main part of this paper.
In contrast to the general theory above
let now $\LG$ be a {\em compact} Lie group throughout this section.
The goal of this section is the classification of the generalized 
connections by the type of their $\Gb$-orbits.
In contrast to the theory of classical connections
in principal fiber bundles, topological subtleties do not play 
an important r\^ole -- a generalized connection is only
an (algebraic) homomorphism from the groupoid $\Pf$ of paths
into the structure group $\LG$, and the generalized gauge transforms are
simply mappings from $M$ to $\LG$. Thus, also the
theory of generalized gauge orbits is governed completely by the
algebraic structure of the action of $\Gb$ on $\Ab$:
\znumgl{
\fbox{ $h_{\qa\circ\qg} (\gamma) = g_x^{-1} \: h_\qa(\gamma) \: g_y$ }
\text{\hspace{2em} for all $\qa\in\Ab$, $\qg\in\Gb$, $\gamma\in\pf{xy}$.}}
For each element $\qg$ of the stabilizer $\bz(\qa)$ of a 
connection $\qa$ the following must be fulfilled:
\znumgl{\label{bedingstabqa}
h_\qa(\gamma) = h_{\qa\circ\qg} (\gamma) = g_x^{-1} \: h_\qa(\gamma) \: g_y
\text{\hspace{2em} for all $\gamma\in\pf{xy}$,}}
hence, in particular,
\bunum
\item
$h_\qa(\alpha) = g_m^{-1} \: h_\qa(\alpha) \: g_m$ for all 
$\alpha\in\hg\ident\pf{mm}$ and
\item
$h_\qa(\gamma_x) = g_m^{-1} \: h_\qa(\gamma_x) \: g_x$ for all
$x\in M$, whereas $\gamma_x$ is for any $x$ 
some fixed path from $m$ to $x$.
\eunum
Thanks any path $\gamma\in\pf{xy}$ can be written as
$\gamma_x^{-1}\:(\gamma_x\gamma\gamma_y^{-1})\: \gamma_y$, i.e. as
a product of paths in $\hg$ and $\{\gamma_x\}$, both conditions 
are even equivalent to \eqref{bedingstabqa}.
From the first condition follows that $g_m$ has to commute with all
holonomies $h_\qa(\alpha)$, i.e. $g_m$ is contained in the centralizer 
$Z(\holgr_\qa)$ of the holonomy group of $\qa$.
Writing the second condition as 
\znumgl{
\label{abbzholgrGb}
g_x = h_\qa(\gamma_x)^{-1} \: g_m \: h_\qa(\gamma_x)
\text{\hspace{2em} for all $x\in M$},}
we see that an element $\qg$ of the stabilizer of $\qa$ is already
completely determined by its value in the point $m$, i.e. by an element of
the holonomy centralizer
$Z(\holgr_\qa)$. From this the isomorphy 
of $\bz(\qa)$ and $Z(\holgr_\qa)$ follows immediately. 

Due to general theorems of the theory of transformation groups 
the gauge orbit $\qa\circ\Gb$ is homeomorphic to the factor space $\nkla$. 
Since $\bz(\qa)$ and $Z(\holgr_\qa) \iso Z(\holgr_\qa) \kreuz \{e_{\Gb_0}\}$ 
are homeomorphic,\footnote{The subgroup $\Gb_0\teilmenge\Gb$
is defined by $\pi_m^{-1}(e_\LG)$. This means, it contains all 
gauge transforms that are trivial in $m$. Obviously, we have
$\Gb\iso\LG\kreuz\Gb_0$.}
we get for the moment {\em heuristically}\/
\zgl{\nkla 
      \iso \rnkl{\Bigl(Z(\holgr_\qa) \kreuz \{e_{\Gb_0}\}\Bigr)}
                {\Bigl(\LG\kreuz\Gb_0\Bigr)}
      \iso \Bigl(\nklza\Bigr) \kreuz \Gb_0.}
We will prove that the left and the right space are indeed 
homeomorphic, i.e. the homeomorphism type of a 
gauge orbit is already determined by that of $\nklza$.
Consequently, two connections have homeomorphic gauge orbits, in particular,
if the holonomy centralizers are conjugate. 

Finally, we can prove that the stabilizers of two
connections are conjugate w.r.t. $\Gb$ 
iff the corresponding holonomy centralizers 
are conjugate w.r.t. $\LG$.
This allows us to define the type of a connection not only (as known from
the general theory of transformation groups) by the
$\Gb$-conjugacy class of its stabilizer
$\bz(\qa)$, but equivalently by the $\LG$-conjugacy class
of its holonomy centralizer $Z(\holgr_\qa)$.

After all, we again mention that in the following $\LG$ is a compact
Lie group. The purely algebraic results, of course, are valid also
without this assumption.

\subsection{Stabilizer of a Connection}
\bdf
Let be $\qa\in\Ab$. Then 
$\eo_\qa:=\qa\circ\Gb\ident\{\qa'\in\Ab\mid\esex\qg\in\Gb:\qa' = \qa\circ\qg\}$
is called \df{gauge orbit} of $\qa$.
\edf
Obviously, two gauge orbits are equal or disjoint.

We need some notations.
\bdf
\label{defholgrbz}
Let $\qa\in\Ab$ be given.
\bnum{3}
\item
The \df{holonomy group} $\holgr_\qa$ of $\qa$ is equal to
$h_\qa(\hg)\teilmenge\LG$.
\item
The centralizer $Z(\holgr_\qa)$ of the holonomy group, also called 
\df{holonomy centralizer} of $\qa$, is the set 
of all elements in $\LG$ that commute with all elements in
$\holgr_\qa$.
\item
The \df{base centralizer} $\bz(\qa)$ of 
$\qa$ is the set of all elements $\qg = (g_x)_{x\in M}$ in $\Gb$ such that 
$h_\qa(\gamma) = g_m^{-1} \: h_\qa(\gamma) \: g_x$
for all $x\in M$ and all paths $\gamma$ from $m$ to $x$.
\enum
\edf
Note that for regular connections the holonomy group defined above 
is exactly the holonomy group known from the classical theory.
We get immediately from the definitions 
\blem
\label{bzugru}
Let be $\qa\in\Ab$ and $\qg\in\Gb$.
\bnum{5}
\item
The holonomy group $\holgr_\qa$ is a subgroup of $\LG$.
\item
$Z(\holgr_\qa)$ is a closed subgroup of $\LG$.
\item
\label{eichtrfhologr}
We have $\holgr_{\qa\circ\qg} = g_m^{-1} \: \holgr_{\qa} \: g_m$ and
$Z(\holgr_{\qa\circ\qg}) = g_m^{-1} \: Z(\holgr_\qa) \: g_m$.
\item
\label{bfqaalleexist}
We have $\qg\in\bz(\qa)$ iff
\bnum{2}
\item
$g_m \in Z(\holgr_\qa)$ and
\item
for all $x\in M$
there is a path $\gamma$ from $m$ to $x$ with
$h_\qa(\gamma) = g_m^{-1} \: h_\qa(\gamma) \: g_x$.
\enum
\enum
\elem
\bpf
\bnum{3}
\item
This is an obvious consequence of homomorphy property of $h_\qa:\hg\nach\LG$.
\item
Trivial.
\item
This follows immediately from
$h_{\qa\circ\qg}(\alpha) = g_m^{-1} h_\qa(\alpha) g_m$ for all $\alpha\in\hg$.
\item
\bhin
We have to prove only $g_m\in Z(\holgr_\qa)$, but this is clear because
we have $h_\qa(\alpha) = g_m^{-1} h_\qa(\alpha) g_m$ for all
$\alpha\in\hg$ by assumption.
\ehin
\brueck
Let $x\in M$ be fixed and $\delta$ be an arbitrary path from $m$ to $x$. Choose a
$\gamma$ such that $h_\qa(\gamma) = g^{-1}_m \: h_\qa(\gamma) \: g_x$. Then
$\alpha:=\delta\gamma^{-1}\in\hg$ and
\berlgl
g^{-1}_m \: h_\qa(\delta) \: g_x
   & = & g^{-1}_m \: h_\qa(\alpha\gamma) \: g_x \\
   & = & g^{-1}_m \: h_\qa(\alpha) \: h_\qa(\gamma) \: g_x \\
   & = & h_\qa(\alpha) \: g^{-1}_m \: h_\qa(\gamma) \: g_x 
         \erl{since $g_m\in Z(\holgr_\qa)$}\\
   & = & h_\qa(\alpha) \: h_\qa(\gamma) 
         \erl{by the choice of $\gamma$}\\
   & = & h_\qa(\delta).
\eerlgl
\qed
\erueck
\enum
\epf
Now we can determine the stabilizer of a connection.
\bprop
\label{kritinstab}
For all $\qa\in\Ab$ and all $\qg\in\Gb$ we have
\zgl{\qa\circ\qg = \qa \aequ \qg\in\bz(\qa).}
\eprop
\bpf
Per def. we have
\zglnum{\qa \circ \qg = \qa 
  \aequ \forall x,y\in M, \gamma\in\pf{xy}: 
        h_\qa(\gamma) = h_{\qa \circ \qg}(\gamma) =
        g_x^{-1} \: h_\qa(\gamma) \: g_y.}{stabbed}

\bhin
Let $\qa\circ\qg = \qa$. 
Due to \eqref{stabbed} $g_m^{-1} \: h_\qa(\alpha) \: g_m = h_\qa(\alpha)$
holds for all $\alpha\in\pf{mm}\ident\hg$, i.e. $g_m\in Z(\holgr_\qa)$.
Again by \eqref{stabbed} we have 
$h_\qa(\gamma_x) = g_m^{-1} \: h_\qa(\gamma_x) \: g_x$
for all $x\in M$ and all $\gamma\in\pf{mx}$. Thus, $\qg\in\bz(\qa)$.
\ehin
\brueck
Let $\qg\in\bz(\qa)$ and $x,y\in M$ be given. Choose some
$\gamma_x\in\pf{mx}, \gamma_y\in\pf{my}$. Then for all $\gamma\in\pf{xy}$
the following holds:
\berlgl
g_x^{-1} \: h_\qa(\gamma) g_y
 & = & g_x^{-1} \: h_\qa(\gamma_x^{-1} \gamma_x \gamma \gamma_y^{-1} \gamma_y) 
       \: g_y \\
 & = & g_x^{-1} \: h_\qa(\gamma_x^{-1}) \: g_m \:\: 
       g_m^{-1} \: h_\qa(\gamma_x \gamma \gamma_y^{-1}) \: g_m \:\: 
       g_m^{-1} \: h_\qa(\gamma_y) \: g_y \\
 & = & (g_m^{-1} \: h_\qa(\gamma_x) \: g_x)^{-1} \:\: 
       h_\qa(\gamma_x \gamma \gamma_y^{-1}) \:\: 
       (g_m^{-1} \: h_\qa(\gamma_y) \: g_y) \\
 &   & \erl{since $\gamma_x \gamma \gamma_y^{-1} \in \hg$ 
            and $g_m\in Z(\holgr_\qa)$}\\
 & = & h_\qa(\gamma_x)^{-1} \:\: 
       h_\qa(\gamma_x \gamma \gamma_y^{-1}) \:\: 
       h_\qa(\gamma_y) \\
 & = & h_\qa(\gamma).
\eerlgl
By \eqref{stabbed} we have $\qa \circ \qg = \qa$.
\qed
\erueck
\epf
Since for compact transformation groups every stabilizer is
closed (see, e.g., \cite{Bredon}), we have using the proposition above
\bcorr
$\bz(\qa)$ is a closed, hence compact subgroup of $\Gb$.
\ecorr
Furthermore, by the lemma above we get
$\qa \circ \qg_1 = \qa \circ \qg_2 
\aequ \qa \circ \qg_1 \circ \qg_2^{-1} = \qa
\aequ \qg_1 \circ \qg_2^{-1} \in \bz(\qa)$, i.e. we can identify
$\eo_\qa$ and $\nkla$ by
\fktdefabgesetzt{\tau}{\nkla}{\eo_\qa.}{[\qg]}{\qa\circ\qg}
Again by the general theory of compact transformation groups 
we get \cite{Bredon}
\bprop
\label{eoiso}
$\tau:\nkla\nach\eo_\qa$ is an equivariant isomorphism between
compact Hausdorff spaces.
\eprop

\subsection{Isomorphy of $\bz(\qa)$ and $Z(\holgr_\qa)$}
In the next subsection we shall 
determine the homeomorphism class of a gauge orbit $\eo_\qa$.
For that purpose, we should use the base centralizer. But, this 
object seems -- at least for the first moment -- to be 
quite inaccessible from the algebraic point of view.
However, looking carefully at its definition (Def. \ref{defholgrbz})
one sees that for given $\qa$ 
due to $h_\qa(\gamma) = g_m^{-1} h_\qa(\gamma) g_x$ 
the value of $g_x$ is already determined by $g_m\in Z(\holgr_\qa)$. 
Therefore, the base centralizer is completely determined by the 
holonomy centralizer.
\bprop
\label{algbz}
For any $\qa\in\Ab$ the map
\fktdefabgesetzt{\phi}{\bz(\qa)}{Z(\holgr_\qa)}{\qg}{g_m} 
is an isomorphism of Lie groups. 

(The topologies on $\bz(\qa)$ and $Z(\holgr_\qa)$
are the relative ones induced by $\Gb$ and $\LG$, respectively.)
\eprop
\bpf
\bunum
\item
Obviously, $\phi$ is a homomorphism. 
\item
Surjectivity

Let $g\in Z(\holgr_\qa)$. Choose for each $x\in M$ a path $\gamma_x$ from
$m$ to $x$ (w.l.o.g. $\gamma_m$ is the trivial path) and define
\zglnum{g_x := h_\qa(\gamma_x)^{-1} \: g \: h_\qa(\gamma_x).}{isobzZholgr}
Obviously, $\qg = (g_x) \in\Gb$ and $\phi(\qg) = g$. 
By Lemma \ref{bzugru}, \ref{bfqaalleexist}
we have $\qg\in\bz(\qa)$ because
\bnum{2}
\item
$g_m = h_\qa(\gamma_m)^{-1} \: g \: h_\qa(\gamma_m) = g \in Z(\holgr_\qa)$
by the triviality of $\gamma_m\in\hg$ and
\item
$h_\qa(\gamma_x) = g_m^{-1} \: h_\qa(\gamma_x) \: g_x$ for the $\gamma_x$
chosen above.
\enum
\item
Injectivity

Clear, because $g_x$ is uniquely determined by $\qa$ and so $g_m$ is
due to $h_\qa(\gamma_x) = g_m^{-1} \: h_\qa(\gamma_x) \: g_x$.
\item
Continuity of $\phi$

$\phi$ is the restriction of $\pi_m : \Gb \nach \Gb_m\ident\LG$
to $\bz(\qa)$. The continuity of $\phi$ is now a consequence of the continuity
of $\pi_m$.
\item
Continuity of $\phi^{-1}$

$\phi:\bz(\qa)\nach Z(\holgr_\qa)$ is a continuous and bijective map
of a compact space onto a Hausdorff space. Therefore,
$\phi^{-1}$ is continuous.
\qed
\eunum
\epf
Finally, we note that obviously the isomorphism $\phi$ does not depend on
the special choice of the paths $\gamma_x$.
\subsection{Determination of the Homeomorphism Class}
As we have seen in the last subsection
$\bz(\qa)$ and $Z(\holgr_\qa)\kreuz\{e_{\Gb_0}\}$ are homeomorphic 
subgroups of $\Gb$. One could conjecture that consequently
\zgl{\text{$\nkla$ and 
$\rnkl{\Bigl(Z(\holgr_\qa) \kreuz \{e_{\Gb_0}\}\Bigr)}
      {\Bigl(\LG\kreuz\Gb_0\Bigr)}
 \iso \Bigl(\nklza\Bigr) \kreuz \Gb_0$}}
are homeomorphic. But, this is not clear at all. For instance, 
$2\Z$ and $3\Z$ are isomorphic, but $\Z/2\Z=\{0,1\}$ and $\Z/3\Z=\{0,1,2\}$ 
are not. Nevertheless, in our case the claimed relation holds:
\bprop
\label{stratabdllem3}
For any $\qa\in\Ab$ there is a homeomorphism 
\zgl{\redisoeo:\Gb_0\kreuz\nklza\nach\nkla.}
\eprop 
Hence, the homeomorphism type of $\eo_\qa$ is not only determined by 
$\nkla$, but already by $\nklza$.

Before we will prove this proposition, 
we shall motivate our choice of the homeomorphism.
First we again choose for each $x\in M$ a path $\gamma_x$ from
$m$ to $x$ where w.l.o.g. $\gamma_m$ is the trivial path.
By equation \eqref{isobzZholgr} we get a homomorphism
\fktdefabgesetzt{\phi'}{\LG}{\Gb}
            {g}{\bigl(h_\qa(\gamma_x)^{-1}\:g\:h_\qa(\gamma_x)\bigr)_{x\in M}}
with $\phi'(Z(\holgr_\qa)) = \bz(\qa)$ and therefore
a map from $\nklza$ to $\nkla$. Furthermore, we have
$\phi'(\LG)\:\Gb_0 = \Gb \iso \phi'(\LG)\kreuz\Gb_0$ with
$\qg\auf \bigl(\phi'(g_m),\phi'(g_m)^{-1}\qg\bigr)$.
Although there is no group structure on $\nkla$ --
in general, $\bz(\qa)$ is only a subgroup and not a normal subgroup of $\Gb$ --,
there is at least a canonical right action of $\Gb$ and $\Gb_0$, respectively,
by $[\qg] \circ \qg' := [\qg\:\qg']$.
Thus, $(\qg,[g]) \auf [\phi'(g)]\circ\qg$ is a good candidate to become 
our desired homeomorphism.
\bpf
First we choose some path $\gamma_x$ from $m$ to $x$ for each $x\in M$
where w.l.o.g. $\gamma_m$ is trivial. Now we define 
\fktdefabgesetzt{\redisoeo}{\Gb_0\kreuz\nklza}{\nkla}
                        {\bigl((g_x)_{x\in M},[g]\bigr)}
                        {\bigl[\phi'(g) \: (g_x)_{x\in M}\bigr]}
with $g_m = e_\LG$.
\bnum{6}
\item
$\redisoeo$ is well-defined.

Let $g_1\aeqrel g_2$, i.e. $g_1 = z g_2$ for some $z\in Z(\holgr_\qa)$.
Define $\qg := (g_x)_{x\in M}\in\Gb_0$. Then we have
\bgl
\redisoeo\bigl((g_x)_{x\in M},[g_1]\bigr) 
 & = & \bigl[\phi'(g_1) \: \qg\bigr] \\
 & = & \bigl[\phi'(z g_2) \: \qg\bigr] \\
 & = & \bigl[\phi'(z) \: \phi'(g_2) \: \qg\bigr] 
       \erl{Homomorphy property of $\phi'$}\\
 & = & \bigl[\phi'(g_2) \: \qg\bigr] 
       \erl{$\phi'(Z(\holgr_\qa)) = \bz(\qa)$ by Proposition \ref{algbz}}\\
 & = & \redisoeo\bigl((g_x)_{x\in M},[g_2]\bigr). 
\egl
\item
$\redisoeo$ is injective.

Let  
$\redisoeo\bigl((g_{1,x})_{x\in M},[g_1]\bigr) = 
 \redisoeo\bigl((g_{2,x})_{x\in M},[g_2]\bigr)$.
Then there exists a $\quer z\in\bz(\qa)$ with
$\phi'(g_1)_x \: g_{1,x} = z_x \: \phi'(g_2)_x \: g_{2,x}$, i.e.
\zgl{
  h_\qa(\gamma_x)^{-1} \: g_1 \: h_\qa(\gamma_x) \: g_{1,x} 
   = z_x \:\: h_\qa(\gamma_x)^{-1} \: g_2 \: h_\qa(\gamma_x) \: g_{2,x} 
}
for all $x\in M$.
Thus, 
\bunum
\item
for $x=m$: $g_1 = z_m g_2$, i.e. $[g_1] = [g_2]$, and
\item
for $x\neq m$:
\bgl
g_{1,x} 
    & = & h_\qa(\gamma_x)^{-1} \: g_1^{-1} \: h_\qa(\gamma_x) \: 
          z_x \: 
          h_\qa(\gamma_x)^{-1} \: g_2 \: h_\qa(\gamma_x) \: g_{2,x} \\
    & = & h_\qa(\gamma_x)^{-1} \: g_1^{-1} \: 
          z_m \:
          g_2 \: h_\qa(\gamma_x) \: g_{2,x} \\
    & = & h_\qa(\gamma_x)^{-1} \: h_\qa(\gamma_x) \: g_{2,x} \\
    & = & g_{2,x},
\egl
\eunum
i.e. $\redisoeo$ is injective.
\bitteseitenumbr
\item
$\redisoeo$ is surjective.

Let $[\wg]\in\nkla$ be given. Define
$g_x := (\phi'(\wg_m)^{-1} \wg)_x$
for all $x\in M$. Then we have
$\redisoeo\bigl((g_x)_{x\in M},[\wg_m]\bigr) = [\wg]$.
\item
$\redisoeo^{-1}$ is continuous.

It is sufficient to prove that the projections $\pr_i\circ\redisoeo^{-1}$
of $\redisoeo^{-1}$ to the factors $\Gb_0$ ($i=1$) 
and $\nklza$ $(i=2)$ are continuous. 
\bnum{2}
\item
$\pr_1\circ\redisoeo^{-1}$ is continuous.

For all $x\in M\setminus\{m\}$ the map

$\begin{array}[t]{cccccc}%
       & \Gb & \txtnach{\pi_{mx}} & \LG\kreuz\LG 
             & \txtnach{\text{mult.}} & \LG\\    
       & \qg & \txtauf{\pi_{mx}} & (g_m, g_x) 
             & \txtauf{\text{mult.}} 
             & (h_{\qa}(\gamma_x)^{-1} g_m^{-1} h_{\qa}(\gamma_x) g_x)
\end{array}$

is a composition of continuous maps and consequently continuous itself.
Since $\pi_{\bz(\qa)}:\Gb\nach\nkla$ is open and surjective, we get
the continuity of $\pi_x\circ\pr_1\circ\redisoeo^{-1}$ for all 
$x\in M\setminus\{m\}$ by 
$\pi_x\circ(\pr_1\circ\redisoeo^{-1})\circ\pi_{\bz(\qa)} = 
     \text{mult.} \circ \pi_{mx}$.
For $x=m$ the statement is trivial.
Thus, $\pr_1\circ\redisoeo^{-1}$ is continuous.
\item
$\pr_2\circ\redisoeo^{-1}$ is continuous.

We use  
$\pi_{Z(\holgr_\qa)} \circ \pi_m 
  = (\pr_2\circ\redisoeo^{-1}) \circ \pi_{\bz(\qa)} : \Gb \nach \nklza$.
The statement now follows because $\pi_{\bz(\qa)}$ is an open and surjective
map and $\pi_{Z(\holgr_\qa)}$ and $\pi_m$ are continuous.  
\enum  
\item
$\redisoeo$ is a homeomorphism
because $\redisoeo^{-1}$ is continuous and bijective.
\qed
\enum
\epf
Thus we get the following important result:
The homeomorphism class of a gauge orbit of a connection is 
completely determined by its holonomy centralizer.
Finally, we should emphasize that, in general, the homeomorphism
$\redisoeo$ is not an equivariant map w.r.t. the
canonical action of $\Gb$ on $\Gb_0\kreuz\nklza$.
\subsection{Criteria for the Homeomorphy of Gauge Orbits}
It is well known that orbits of general transformation groups
are classified by the conjugacy classes of their stabilizers. 
This would effect in our case that the gauge orbits are characterized
by the conjugacy class of their corresponding base centralizer
w.r.t. $\Gb$.
As we have already seen, the base centralizer of a connection
$\qa$ is isomorphic to the holonomy centralizer of $\qa$ and 
the homeomorphism type of the gauge orbit is completely determined
by that of $\nklza$. Now we are going to show that base centralizers 
are conjugate w.r.t. $\Gb$ if and only if the corresponding  
holonomy centralizers are conjugate w.r.t. $\LG$. This will allow us 
to define the type of a gauge orbit $\eo_\qa$ to be the conjugacy class of
$Z(\holgr_\qa)$ w.r.t. $\LG$. The investigation of the set 
of all these classes is much easier than in the case of classes in $\Gb$.

We want to prove the following
\bprop
\label{satz:konj_zentr=konj_bz}
Let $\qa_1,\qa_2\in\Ab$ be two generalized connections. 
Then the following statements are equivalent:
\bnum{2}
\item
$Z(\holgr_{\qa_1})$ and $Z(\holgr_{\qa_2})$ are conjugate in $\LG$.
\item
$\bz(\qa_1)$ and $\bz(\qa_2)$ are conjugate in $\Gb$.
\enum
\eprop
It would be quite easy to prove this directly using Proposition \ref{algbz}. 
Nevertheless, we do not want to do this. Instead, we shall first
derive some concrete criteria for the homeomorphy of two 
gauge orbits. Finally, the just claimed proposition will be a nice by-product.
\bprop
\label{isobzimplisoeo}
Let $\qa_1,\qa_2\in\Ab$ be two generalized connections. Furthermore,
let there exist an isomorphism $\Psi: \Gb \nach \Gb$ of topological groups with 
$\Psi\bigl(\bz(\qa_1)\bigr) = \bz(\qa_2)$.

Then the map
\fktdefabgesetzt{\Phi}{\eo_{\qa_1}}{\eo_{\qa_2}}
                      {\qa_1\circ\qg}{\qa_2\circ\Psi(\qg)}
is a homeomorphism compatible with the action of $\Gb$.
\eprop
\bpf
\bunum
\item
$\Phi$ is well-defined.

Let $\qa_1 \circ \qg = \qa_1 \circ \qg'$. Then we have
$\qa_1 \circ (\qg \circ {\qg'}^{-1}) = \qa_1$, i.e.
$\qg \circ {\qg'}^{-1}\in\bz(\qa_1)$ by Proposition \ref{kritinstab}. By assumption
we have
$\Psi(\qg \circ {\qg'}^{-1}) =
 \Psi(\qg) \circ \Psi({\qg'})^{-1} \in\bz(\qa_2)$, i.e.
$\qa_2 \circ \Psi(\qg) = \qa_2 \circ \Psi(\qg')$.
\item
Since $\Psi$ is a group isomorphism, $\Phi$ is again an isomorphism that is 
compatible with the action of $\Gb$.
\item
For the proof of the homeomorphy property of $\Phi$ we consider the following
commutative diagram:
\begin{center}
\vspace*{\CDgap}
\begin{minipage}{5cm}
\begin{diagram}[labelstyle=\scriptstyle,height=\CDhoehe,l>=3em]
\eo_{\qa_1}            & \rnach^{\Phi}   & \eo_{\qa_2}            \\
\dnach^{\tau_1}_{\iso} &                 & \dnach^{\iso}_{\tau_2} \\
\rnkl{\bz(\qa_1)}{\Gb} & \rnach^{\Omega} & \rnkl{\bz(\qa_2)}{\Gb}
\end{diagram}
\end{minipage}
\hspace*{3em}\begin{minipage}{6cm}
\begin{diagram}[labelstyle=\scriptstyle,height=\CDhoehe,l>=3em]
\qa_1\circ\qg  & \rauf^{\Phi}   & \qa_2\circ\qg    \\
\dauf^{\tau_1} &                & \dauf_{\tau_2}   \\
\phantom{{}_{\bz(\qa_1)}}[\qg]_{\bz(\qa_1)}
               & \rauf^{\Omega} & 
                       [\Psi(\qg)]_{\bz(\qa_2)}\cursorzurueck{{}_{\bz(\qa_2)}}
\end{diagram}
\end{minipage}.
\end{center}
Since $\tau_1$ and $\tau_2$ are homeomorphisms, it is sufficient to prove
the homeomorphy property for $\Omega$.
\bunum
\item
$\Omega$ is well-defined and
bijective due to $\Omega = \tau_2 \circ \Phi \circ \tau_1^{-1}$.
\item
$\Omega$ is continuous.

The map $\pi_{\bz(\qa)} : \Gb \nach \nkla$ is an orbit space projection
for all $\qa \in\Ab$ and consequently surjective, continuous and open. 
Using
$\Omega\circ\pi_{\bz(\qa_1)}=\pi_{\bz(\qa_2)}\circ\Psi$ we see for any 
open $U\teilmenge\rnkl{\bz(\qa_2)}{\Gb}$ that
$\Omega^{-1}(U) = \pi_{\bz(\qa_1)} (\Psi^{-1} (\pi_{\bz(\qa_2)}^{-1}(U)))
\teilmenge \rnkl{\bz(\qa_1)}{\Gb}$ is again open.
\item
$\Omega^{-1}$ is continuous by the same reason as above.
\eunum
Thus, $\Omega$ is a homeomorphism.
\qed
\eunum
\epf
To simplify the speech in the following we state
\bdf
Let $\LG$ be a Lie group (topological group) and let $U_1$ and $U_2$ 
be closed subgroups of $\LG$.

$U_1$ and $U_2$ are called \df{extendibly isomorphic (w.r.t. $\LG$)} iff there
is an isomorphism $\psi:\LG\nach\LG$ of Lie groups (topological groups)
with $\psi(U_1) = U_2$.\footnote{If misunderstanding seems to be unlikely,
we simply drop "w.r.t. $\LG$" and write "extendibly isomorphic".}
\edf

In Proposition \ref{isobzimplisoeo} 
we compared gauge orbits w.r.t. their base centralizers.
Now we will compare them using 
their holonomy centralizers. In order to manage this we need
an extendibility lemma.

Let the holonomy centralizers of two connections
be extendibly isomorphic, i.e. let there exist a  
$\psi:\LG\nach\LG$ with $\psi(Z(\holgr_{\qa_1})) = Z(\holgr_{\qa_2})$.
By $\Psi:=\phi_2^{-1}\circ\psi\circ\phi_1$ the 
base centralizers are isomorphic. Extending $\Psi$ to $\Gb$ we get
\blem
\label{isozholgrimplisobz}
Let $\qa_1,\qa_2\in\Ab$ be two generalized connections. 
Then the following statement holds:

If $Z(\holgr_{\qa_1})$ and $Z(\holgr_{\qa_2})$ are extendibly isomorphic, then
$\bz(\qa_1)$ and $\bz(\qa_2)$ are also extendibly isomorphic.

We have explicitly: Let  
$\psi: \LG \nach \LG$ be an isomorphism of Lie groups with  
$\psi\bigl(Z(\holgr_{\qa_1})\bigr) = Z(\holgr_{\qa_2})$.
Furthermore, let $\gamma_x$ be an arbitrary, but fixed path in $M$ for 
each $x\in M$. Then we have:
\bunum
\item
The map $\Psi : \Gb \nach \Gb$ defined by
\zglnum{
 \Psi(\qg)_x := h_{\qa_2}(\gamma_x)^{-1} \:\:
                \psi\bigl(h_{\qa_1}(\gamma_x)
                          g_x            
                          h_{\qa_1}(\gamma_x)^{-1}\bigr)\:\:
                h_{\qa_2}(\gamma_x)
}{Psi}
is an isomorphism of topological groups. 
\item
$\Psi\einschr{\bz(\qa_1)}$ is an isomorphism of Lie groups between
$\bz(\qa_1)$ and $\bz(\qa_2)$. Furthermore,
$\Psi\einschr{\bz(\qa_1)}$ is independent of the choice of the paths
$\gamma_x$.
\eunum
\elem
\bpf
Let $Z(\holgr_{\qa_1})$ and $Z(\holgr_{\qa_2})$ be
extendibly isomorphic with the corresponding isomorphism $\psi$.
\bunum
\item
Obviously, we have $\Psi(\qg)\in\Gb$ and $\Psi$ is a homomorphism of groups.
Moreover, $\Psi$ is bijective with the inverse
\zglnum{
 \Psi^{-1}(\qg)_x = h_{\qa_1}(\gamma_x)^{-1} \:\: 
                    \psi^{-1}\bigl(h_{\qa_2}(\gamma_x) 
                                   g_x          
                                   h_{\qa_2}(\gamma_x)^{-1}\bigr)\:\:
                    h_{\qa_1}(\gamma_x).
}{Psiinvers}
To prove the continuity of $\Psi$ it is sufficient to prove the continuity
of $\pi_x\circ\Psi$ for all $x$.
Hence, let $U\teilmenge\LG$ be open. Then we have
\bgl
 &   & (\pi_x\circ\Psi)^{-1} (U) \\
 & = & \{\qg\in\Gb\mid (\pi_x\circ\Psi)(\qg) = 
                        \Psi(\qg)_x = \\
 &   & \phantom{\{\qg\in\Gb\mid }\:\:
                               h_{\qa_2}(\gamma_x)^{-1} \:
                               \psi\bigl(h_{\qa_1}(\gamma_x)
                               g_x            
                               h_{\qa_1}(\gamma_x)^{-1}\bigr)\:
                               h_{\qa_2}(\gamma_x) \in U\} \\
 & = & \pi_x^{-1}\bigl(h_{\qa_1}(\gamma_x)^{-1} \: 
                       \psi^{-1}(h_{\qa_2}(\gamma_x)) \:
                        \psi^{-1}(U) \:
                        \psi^{-1}(h_{\qa_2}(\gamma_x)^{-1})\:
                       h_{\qa_1}(\gamma_x)\bigr).
\egl
Since $\psi$ is a homeomorphism and $\pi_x$ is continuous, 
$(\pi_x\circ\Psi)^{-1} (U)$ is open.

The continuity of $\Psi$ is now a consequence of Corollary 
\ref{folg:stetkritGbx}, that of $\Psi^{-1}$ is clear.
\item
Let $\phi_i$ be the isomorphism for $\qa_i$ ($i=1,2$) corresponding to 
Proposition \ref{algbz}. Then we have 
\zgl{\Psi\einschr{\bz(\qa_1)} = \phi_2^{-1} \circ \psi \circ \phi_1 : 
           \bz(\qa_1) \nach \bz(\qa_2).}
Since $\phi_1, \phi_2$ and $\psi$ are Lie isomorphisms and, moreover, 
independent of the choice of the $\gamma_x$,
$\Psi\einschr{\bz(\qa_1)}$ is again an isomorphism of Lie groups that is
independent of the choice of the $\gamma_x$.
\eunum
Thus, $\bz(\qa_1)$ and $\bz(\qa_2)$ are extendibly isomorphic.
\qed
\epf
The next lemma is obvious.
\blem
Let $\qa_1,\qa_2\in\Ab$ be two generalized connections. 
Then $Z(\holgr_{\qa_1})$ and $Z(\holgr_{\qa_2})$ are
extendibly isomorphic provided they are conjugate w.r.t. $\LG$.
\elem
Now we can prove Proposition \ref{satz:konj_zentr=konj_bz}.
\bpf[Proposition \ref{satz:konj_zentr=konj_bz}]
\bunum
\item
Let $Z(\holgr_{\qa_1})$ and $Z(\holgr_{\qa_2})$ be conjugate and thus 
also extendibly isomorphic.
The map $\Psi:\Gb\nach\Gb$ from Lemma \ref{isozholgrimplisobz} fulfills
now 
\zgl{\Psi(\qg') = \bigl( (h_{\qa_1}(\gamma_x)^{-1} \: g \: 
                           h_{\qa_2}(\gamma_x))^{-1} \: g'_x \:
                          (h_{\qa_1}(\gamma_x)^{-1} \: g \: 
                           h_{\qa_2}(\gamma_x))
                           \bigr)_{x\in M}}
where $g\in\LG$ was chosen such that 
$Z(\holgr_{\qa_2}) = (\Ad\: g) Z(\holgr_{\qa_1})$.
We define 
$\qg :=
  \bigl(h_{\qa_1}(\gamma_x)^{-1} \: g \: h_{\qa_2}(\gamma_x)\bigr)_{x\in M}$.
Hence, the map $\Psi:\Gb\nach\Gb$ from Lemma \ref{isozholgrimplisobz}
is simply $\Ad\: \qg$. Moreover, $\Ad\: \qg$ maps $\bz(\qa_1)$ isomorphically 
onto $\bz(\qa_2)$. Thus, $\bz(\qa_2) = (\Ad\: \qg) \bz(\qa_1)$.
\item
Let $\bz(\qa_1)$ and $\bz(\qa_2)$ be conjugate, i.e. let there exist
a $\qg\in\Gb$ with $\bz(\qa_2) = \qg^{-1} \bz(\qa_1) \qg$. Then we obviously
have $Z(\holgr_{\qa_2}) = g_m^{-1} Z(\holgr_{\qa_1}) g_m$.
\qed
\eunum
\epf
Let us summarize:
\bthm
\label{satzimplkette}
Let $\qa_1,\qa_2\in\Ab$ be two generalized onnections. 
Then the following implication chain holds:
\bgl
         && \text{$\bz(\qa_1)$ and $\bz(\qa_2)$ are conjugate in $\Gb$.} \\
 \aequ   && \text{$Z(\holgr_{\qa_1})$ and $Z(\holgr_{\qa_2})$ are conjugate in $\LG$.} \\
 \impliz && \text{$Z(\holgr_{\qa_1})$ und $Z(\holgr_{\qa_2})$ are
                  extendibly isomorphic.} \\
 \impliz && \text{$\bz(\qa_1)$ und $\bz(\qa_2)$ are extendibly isomorphic.} \\
 \impliz && \text{The gauge orbits $\eo_{\qa_1}$ and $\eo_{\qa_2}$ 
                  are homeomorphic.}
\egl
\ethm
This theorem has an interesting and perhaps a little bit surprising 
consequence: Even {\em after} 
projecting $\Ab$ down to $\AbGb\ident\Hom(\hg,\LG)/\Ad$ 
the complete knowledge about the homeomorphism class
of the corresponding gauge orbit is conserved. Naively one would suggest
that after projecting the total gauge orbit onto one single point this
information should be lost.
But, the homeomorphism class is already determined by giving
the holonomy centralizer, that, the other way round, 
can be, up to a global conjugation,
reconstructed from $[\qa]$.
\bprop
For each $[\qa]\in\AbGb$ the homeomorphism class of the gauge orbit 
corresponding to $[\qa]$ can be reconstructed from $[\qa]$.
\eprop
\section{Discussion or How to Define the Gauge Orbit Type}
If we ignored the usual definition of the type of an orbit in a general
$G$-space, then Theorem \ref{satzimplkette} would open us
several possibilities to define the type of a gauge orbit. If
the type should characterize as "uniquely" as possible the 
homeomorphism class of the gauge orbit, then it would be advisable to 
define the base centralizer modulo extendible isomorphy to be the type.
But, even this choice would not guarantee that two gauge orbits with
different type are in fact non-homeomorphic.
Moreover, the base centralizers as subgroups of $\Gb$ are not so easily
controllable as centralizers in $\LG$ are. Thus, we will take the
holonomy centralizer for the definition. It remains only the question,
whether we should take the centralizer modulo conjugation or modulo
extendible isomorphy. We have to collect conjugate centralizers 
in one type anyway in order to make points of one orbit
be of the same type. (Note, that the holonomy centralizers
of two gauge equivalent connections are generally not equal but 
only conjugate.)

If we now include the general definition of an orbit type into our 
considerations again, it will be clear that we shall use the centralizer modulo
conjugation. 
But, since two connections have one and the same (usual) orbit type
iff their base centralizers are conjugate, i.e. iff
their holonomy centralizers are conjugate,
we define the {\em gauge} orbit type by
\bdf
The \df{type} of a gauge orbit $\eo_\qa$
is the holonomy centralizer of $\qa$ modulo conjugation.
\edf
We emphasize that this definition of the type of the gauge orbit $\eo_\qa$
is -- as mentioned above --
independent of the choice of the connection $\qa'\in\eo_\qa$. In fact, if
$\qa'$ is gauge equivalent to $\qa$, by Lemma \ref{bzugru} there is a
$g\in\LG$ with $Z(\holgr_{\qa'}) = g^{-1} Z(\holgr_\qa) g$. 
Hence, the holonomy centralizers of $\qa$ and $\qa'$ are conjugate.
Thus, we can assign to each $[\qa]\in\AbGb$ a unique gauge orbit type. 

Using Theorem \ref{satzimplkette} we get immediately
\bcorr
\label{gltypimplhomeoeo}
Two gauge orbits with the same type are homeomorphic.
\ecorr
Finally, we want to give a further justification for our definition 
of the gauge orbit type. Let us consider regular connections. 
In the literature there are two different definitions for the type 
of a "classical" gauge orbit: On the one hand \cite{f11}, one 
chooses the total stabilizer of 
$A\in\A$ in $\G$. 
On the other hand \cite{f12},
one sees first that the pointed gauge group $\G_0$ (the set of
all gauge transforms that are the identity on a fixed fibre) is
a normal and closed subgroup in $\G$. Obviously, $G:=\G/\G_0$ can be
identified with the structure group $\LG$. Moreover, 
the action of $\G_0$ on $\A$ is free, proper and smooth.
This way one gets an action of $G$, the "essential part" of the gauge
transforms, on the space $\A/\G_0$. Now, the gauge orbit types are the
conjugacy classes of stabilizers being closed subgroups of $G\iso\LG$.
This definition corresponds to our choice of the 
centralizer of the holonomy group.
Due to the statements proven above these two descriptions are equivalent if we
consider generalized connections, but in general {\em not} if we work in
the classical
framework. There it is under certain circumstances possible \cite{f12} that 
two connections though have conjugate holonomy centralizers, but this 
conjugation cannot be lifted to a conjugation of the 
base centralizers. 
The deeper reason behind this is that the gauge transform 
$\qg = \bigl(h_{A_1}(\gamma_x)^{-1}\: g\: h_{A_2}(\gamma_x)\bigr)_{x\in M}$ 
(cf. proof of Proposition \ref{satz:konj_zentr=konj_bz})
generally is not a classical gauge transform, i.e. it is not smooth.
Nevertheless, in case of the definition using the holonomy group we have
\bcorr
The gauge orbit type is conserved by the embedding $\A\einbett\Ab$.
\ecorr
But, note that this does not mean at all that the classical and the  
generalized gauge orbit of a classical connection itself are equal or 
at least homeomorphic. 

\section{Acknowledgements}
I am grateful to Gerd Rudolph, Matthias Schmidt and 
Eberhard Zeidler guiding me to the
theory of gauge orbits. Additionally, I thank Gerd Rudolph for reading 
the drafts. Moreover, I thank Jerzy Lewandowski for asking me
how the notion of webs is related to the notion of paths in the present paper.
Finally, I thank the Max-Planck-Institut f\"ur Mathematik
in den Naturwissenschaften in Leipzig for its generous support.
\addcontentsline{toc}{section}{Literaturverzeichnis}
\bibliographystyle{plain}

\begin{thebibliography}{10}

\bibitem{a72}
Abhay Ashtekar and C.~J. Isham.
\newblock Representations of the holonomy algebras of gravity and nonabelian
  gauge theories.
\newblock {\em Class. Quant. Grav.}, {\bf 9}:1433--1468, 1992.

\bibitem{a48}
Abhay Ashtekar and Jerzy Lewandowski.
\newblock Representation theory of analytic holonomy {$C^*$} algebras.
\newblock {In {\em Knots and Quantum Gravity}, edited by John C. Baez
          (Oxford Lecture Series in Mathematics and its Applications),
           Oxford University Press, Oxford, 1994.}

\bibitem{a28}
Abhay Ashtekar and Jerzy Lewandowski.
\newblock Differential geometry on the space of connections via graphs and
  projective limits.
\newblock {\em J. Geom. Phys.}, {\bf 17}:191--230, 1995.

\bibitem{a30}
Abhay Ashtekar and Jerzy Lewandowski.
\newblock Projective techniques and functional integration for gauge theories.
\newblock {\em J. Math. Phys.}, {\bf 36}:2170--2191, 1995.

\bibitem{a6}
Abhay Ashtekar, Jerzy Lewandowski, Donald Marolf, Jose Mour{\~a}o, and Thomas
  Thiemann.
\newblock {$SU(N)$} quantum {Y}ang-{M}ills theory in two-dimensions: A complete
  solution.
\newblock {\em J. Math. Phys.}, {\bf 38}:5453--5482, 1997.

\bibitem{d3}
John~C. Baez and Stephen Sawin.
\newblock Functional integration on spaces of connections.
\newblock {\em J. Funct. Anal.}, {\bf 150}:1--26, 1997.

\bibitem{d17}
John~C. Baez and Stephen Sawin.
\newblock Diffeomorphism-invariant spin network states.
\newblock {\em J. Funct. Anal.}, {\bf 158}:253--266, 1998.

\bibitem{Bredon}
Glen~E. Bredon.
\newblock {\em Introduction to Compact Transformation Groups}.
\newblock Academic Press, Inc., New York, 1972.

\bibitem{paper3}
Christian Fleischhack.
\newblock {Hyphs and the Ashtekar-Lewandowski Measure}.
\newblock MIS-Preprint 3/2000, math-ph/0001007.

\bibitem{paper4}
Christian Fleischhack.
\newblock {Stratification of the Generalized Gauge Orbit Space}.
\newblock MIS-Preprint 4/2000, math-ph/0001008.

\bibitem{paper1}
Christian Fleischhack.
\newblock {A new type of loop independence and $SU(N)$ quantum Yang-Mills
  theory in two dimensions}.
\newblock {\em J. Math. Phys.}, {\bf 41}:76--102, 2000.

\bibitem{e14}
R.~Giles.
\newblock Reconstruction of gauge potentials from {W}ilson loops.
\newblock {\em Phys. Rev.}, {\bf D24}:2160--2168, 1981.

\bibitem{Kelley}
John~L. Kelley.
\newblock {\em General Topology}.
\newblock D. van Nostrand Company, Inc., Toronto, New York, London, 1955.

\bibitem{f11}
Witold Kondracki and Jan Rogulski.
\newblock {On the stratification of the orbit space for the action of
  automorphisms on connections (Dissertationes mathematicae 250)}.
\newblock Warszawa, 1985.

\bibitem{f12}
Witold Kondracki and Pawel Sadowski.
\newblock {Geometric structure on the orbit space of gauge connections}.
\newblock {\em J. Geom. Phys.}, {\bf 3}:421--434, 1986.

\bibitem{d20}
Jerzy Lewandowski.
\newblock Group of loops, holonomy maps, path bundle and path connection.
\newblock {\em Class. Quant. Grav.}, {\bf 10}:879--904, 1993.

\bibitem{e46}
Jerzy Lewandowski and Thomas Thiemann.
\newblock Diffeomorphism invariant quantum field theories of connections in
  terms of webs.
\newblock {\em Class. Quant. Grav.}, {\bf 16}:2299--2322, 1999.

\bibitem{a42}
Donald Marolf and Jose~M. Mour{\~a}o.
\newblock On the support of the {A}shtekar-{L}ewandowski measure.
\newblock {\em Commun. Math. Phys.}, {\bf 170}:583--606, 1995.

\bibitem{f1}
P.~K. Mitter.
\newblock Geometry of the space of gauge orbits and the {Y}ang-{M}ills
  dynamical system.
\newblock {Lectures given at Carg\'ese Summer Inst. on Recent
           Developments in Gauge Theories, Carg\`ese, 1979}.

\bibitem{e8}
Alan~D. Rendall.
\newblock Comment on a paper of {A}shtekar and {I}sham.
\newblock {\em Class. Quant. Grav.}, {\bf 10}:605--608, 1993.

\bibitem{f14}
Gerd Rudolph, Matthias Schmidt, and Igor Volobuev.
\newblock {Classification of gauge orbit types for $SU(n)$ gauge theories}
\newblock (in preparation).

\end{thebibliography}

\end{document}